\documentclass{article}

\usepackage[english]{babel}

\usepackage[letterpaper,top=2cm,bottom=2cm,left=3cm,right=3cm,marginparwidth=1.75cm]{geometry}

\usepackage{authblk}
\usepackage[colorlinks=true, allcolors=blue]{hyperref}
\usepackage{graphicx} 
\usepackage{amsmath}
\usepackage{amssymb}
\usepackage{amsthm}
\usepackage{microtype}
\usepackage{booktabs}
\usepackage{xcolor}
\usepackage{mathtools}
\usepackage{enumerate}
\usepackage{tipa}
\usepackage{algorithm2e}
\usepackage{mdframed}
\def\es{exponent-string}
\def\bino{\oplus{}}
\def\rates{$\mathbb{Q}^+$-\es}
\def\reales{$\mathbb{R}^+$-\es}
\def\Resset{$\Sigma_{\mathbb{R}^+}^*$}
\def\Qesset{$\Sigma_{\mathbb{Q}^+}^*$}
\DeclareMathOperator*{\argmin}{arg\,min}
\newtheorem{theorem}{Theorem}
\newtheorem{lemma}{Lemma}
\newtheorem{proposition}{Proposition}
\newtheorem{definition}{Definition}
\newtheorem{remark}{Remark}
\newtheorem{corollary}{Corollary}
\newtheorem{observation}{Observation}
\title{Exponent-Strings and Their Edit Distance}

\begin{document}
\author[]{Ingyu Baek}
\affil[]{
Yonsei University, Seoul, Republic of Korea\\
\textit{ingyubaek@yonsei.ac.kr}
}
\date{}
\maketitle

\begin{abstract}
An \es{} is an extension of traditional strings
that can incorporate real-numbered exponents,
indicating the quantity of characters.
This novel representation overcomes the limitations of traditional discrete string by enabling
precise data representation for applications such as phonetic transcription that contains sound duration.

Although applications of exponent-string are focused on exponent-string with real-numbered exponents, 
formal definition uses arbitrary semigroup.
For any semigroup $S$, $S$-\es{s} are allowed to have
elements of $S$ as exponents.
We investigate algebraic properties of $S$-\es{s} and further justify
that \reales{} is a natural extension of the string. 

Motivated by the problem of calculating the similarity between 
spoken phone sequence and correct phone sequence,
we develop exp-edit distance---a specialized metric
designed to measure the similarity between \reales{s}.
By extending the traditional string edit distance to handle continuous values,
exp-edit distance deals with \reales{s} that embody both discrete and continuous properties.
Our exploration includes a rigorous mathematical formulation of exp-edit distance and an algorithm to compute it.
\end{abstract}
\section{Introduction}
String is defined over a set of symbols or an alphabet, as a finite sequence of the symbols.
For instance, let $\Sigma=\{a,b,c\}$ be an alphabet.
Then, $(a,a,c,c,b,b,b)$ is a valid string, while 
$aaccbbb$ is a notation that is more widely used.
The power of characters may be used to allow even compacter representation: $a^2c^2b^3$.
Same idea of using pairs of symbol and number to represent string, is referred differently as Run-length~encoding~(RLE) 
in the context of the data compression~\cite{quek00}.

We introduce exponent-string that is a continuous generalization of traditional string, by 
allowing non-integer value as an exponent of the characters.
For instance, $a^{1.3}c^{2.1}b^{3.6}$ is a valid exponent-string.

In fields such as phonetic transcription~\cite{Johnson04,LevinsonLM90,Pitman1848},
where precise duration and subtle nuances of sounds~\cite{LadefogedM90} are crucial,
using traditional string that consists of discrete characters may not be enough.
The International Phonetic Alphabet~(IPA)~\cite{Lauf89,OconnellK94}
employs suprasegmentals~\cite{bruce1989report} that represent relative duration~\cite{Thorsen87} rather than absolute values.
This system often fails to capture the actual length of vowel sounds accurately,
which can be vital for applications like speech recognition and linguistic analysis~\cite{KessensS04,PamisettyS23,WesterKCS01}.
Additionally, the variability in vowel duration due to factors like stress, surrounding phonetic environment,
and speaker characteristics~\cite{Docherty92,Ohman66} can significantly affect the accuracy of these transcriptions.
The current transcription systems do not provide standardized symbols for precise duration~\cite{Keating88},
leading to inconsistencies and potential misinterpretations across different linguistic analyses~\cite{SaraclarK00}.

Exponent-strings enrich the traditional string model by incorporating exponents that denote the duration or magnitude of each character explicitly.
This allows a more granular representation of data, accommodating continuous variables directly within the string structure.
An \es{} such as $\mathtt{a^2c^{1.5}b^3}$ quantifies the
respective durations 
of characters \texttt{a}, \texttt{c}, and \texttt{b}
as 2, 1.5 and 3 units.
This extension provides a significant improvement over traditional phonetic transcription
by enabling the representation of both discrete and continuous data attributes.

The capability of \es{s} to denote exact durations allows for more precise computation of similarities and distinctions in phonetic transcriptions.
For example, the differences in pronunciation duration of vowel sounds between words like \texttt{beat}
and \texttt{bead} can be explicitly represented,
which is crucial for distinguishing between similar sounds.
Likewise, there are several word pairs such as \emph{mace-maze} and \emph{fort-fault}~\cite{KluenderDW88,YinZC21}. 
Table~\ref{tab:exponent-example} illustrates how \es{s} can represent these durations more clearly compared to traditional methods.

\begin{table}[hbt]
    \centering
    \begin{tabular}{cc|cc|cc|cc}
    \hline
    \multicolumn{8}{l}{Traditional phonetic transcription examples} \\\hline
    {[bi\textlengthmark{}t]} &{beat}  &{[bi\textlengthmark{}d]} &{bead}
    &{[f\textipa{O}\textlengthmark{}t]} &{fort}  &{[f\textipa{O}\textlengthmark{}lt]} & {fault}\\\hline
    \multicolumn{8}{l}{Exponential-string to denote duration for phonetic transcription} \\\hline
    {[b$^1$i$^{1.9}$t$^1$]} &{beat}  &{[b$^1$i$^{3.5}$d$^1$]} &{bead}
    &{[f$^{1}$\textipa{O}$^{1.7}$t$^{1}$]} &{fort}  &{[f$^{1}$\textipa{O}$^{2.5}$l$^{1}$t$^{1}$]} & {fault}\\\hline
    \end{tabular}
    \caption{An illustration of \es{} as a duration-explicit notation for phonetic transcription
    that overcomes the limitations of current systems.}
    \label{tab:exponent-example}
\end{table}

Beyond enriching string representations, our study also extends into exploring the edit distance for \es{s}.
This is crucial for accurately measuring similarities in phonetic transcription.
One important practical usage is automatic detection of paraphasia~\cite{LeLP17,Smith19}. 
For instance, Le et al.~\cite{LeLP17} consider edit distance
between spoken phone sequence and correct phone sequence,
and guesses whether the speech was paraphasic, 
using classification algorithms such as decision tree.
This extension of string edit distance explores its properties within the context of \es{s}.
The theoretical foundation of \es{s} lies in their ability to bridge discrete string manipulation with continuous mathematical modeling.
This duality offers a robust representation for addressing non-integer values such as duration and magnitude where traditional strings or RLE fall short.
By integrating real-numbered exponents, \es{s} can precisely model phenomena that vary continuously over time or magnitude,
a capability not inherently available in standard string representations or RLE.

In the subsequent sections, we will explore the formal definition of \es{s} and discuss their algebraic properties.
Then we define exp-edit distance, which is an edit distance for exponent-strings. 
The main difference between string edit distance and exp-edit distance is that
exp-edit operations are allowed to edit non-integer amount of the symbol. 
For instance, $a^{0.5}\rightarrow\lambda$ is a valid delete operation that requires
half the cost of ordinary deletion $a\rightarrow\lambda$.
We will provide an algorithm that computes exp-edit distance and prove its correctness.
The proof would contain some of the most important properties of the exp-edit distance.
For instance, Corollary~\ref{cor:string-pseudo-equiv} shows that 
exp-edit distance is an extension of the string edit distance.
Then, we further analyze properties of the exp-edit distance and its relation with 
traditional string edit distance.
\section{Preliminaries}\label{sec:prelim}
\subsection{Mathematical Background}\label{ssec:mathematical-background}
A semigroup is a pair consisting of a set and a binary operation defined over such set, that is 
(1) closed under the operation and (2) the operation satisfies 
associativity. No other conditions are implied.
Monoid is a special kind of semigroup. For a semigroup to be a monoid, the set must contain an identity element.

We say $X=(\mathcal{X},\bino{}_X)$ is a subsemigroup of
semigroup $Y=(\mathcal{Y},\bino{}_Y)$ 
if (1)$\mathcal{X}\subseteq \mathcal{Y}$, 
(2) for $\forall a,b\in\mathcal{X}$, 
$a \bino{}_Y b = a\bino{}_X b \in \mathcal{X}$.
Similarly, 
$X=(\mathcal{X},\bino{}_X)$ is a submonoid of
monoid $Y=(\mathcal{Y},\bino{}_Y)$ 
if (1)$\mathcal{X}\subseteq \mathcal{Y}$, 
(2) for $\forall a,b\in\mathcal{X}$, 
$a \bino{}_Y b = a\bino{}_X b \in \mathcal{X}$
and (3) $\mathcal{X}$ contains an identity element of $\mathcal{Y}$.

Let $\mathbb{N}$, $\mathbb{Q}$, 
$\mathbb{Q}^+$,
$\mathbb{R}$, $\mathbb{R}^+$ and
$\mathbb{R}_{\geq 0}$ denote 
set of natural, rational, positive rational, real,  positive real and nonnegative real numbers, respectively.
In this paper, natural numbers refer to integers greater or equal to 1, not including zero.
When considering semigroup of $\mathbb{N}, \mathbb{Q}^+$ or $\mathbb{R}^+$,
unless specified otherwise, 
we assume addition $+$ is the binary operation that constructs semigroup.

For a set~$A$ and its elements~$x,y\in A$, we define a function~$\delta$ to be 
\begin{align*}
    &\delta_{x}(y)=\delta(x,y)=\begin{cases}
        1, &\textnormal{ if } x=y;\\
        0, &\textnormal{ otherwise.}
    \end{cases}
\end{align*}
\begin{proposition}[Jackson and Agunwamba~\cite{JacksonA77}]\label{prop:glb-property}
    For an arbitrary nonempty subset~$X\subseteq \mathbb{R}$,
    if $X$ has a lower bound in real number,
    then $X$ also has the greatest lower bound or the infimum in real number.
\end{proposition}
Note that  Proposition~\ref{prop:glb-property} does not hold for rational numbers;
for example, $A=\{x|x^2>2,x\in\mathbb{Q}\}$ has a lower bound in rationals 
such as $0, 1$ or $1.414$,
but its greatest lower bound $\sqrt{2}$ is not a rational number.

\subsection{String Edit Distance}
\begin{definition}\label{def:string-edit-distance}
The \emph{string edit distance} from strings~$x$ to $y$ 
is the minimum total cost of basic operations that transform $x$ to $y$.
Here we consider three basic operations: insertion, deletion and
substitution of a single character.
\end{definition}
\begin{definition}\label{def:string-cost-function}
The \emph{cost function} of each operation, insertion, deletion and substitution is denoted as follows:
\begin{enumerate}
\item\label{def:string-cf-insertion} The cost function for insertion is $w_{ins}:\Sigma \rightarrow \mathbb{R}^+$.
$w_{ins}(a)$ of a single character~$a$ can also be denoted as $w(\lambda \rightarrow a)$.
\item\label{def:string-cf-deletion} The cost function for deletion is $w_{del}: \Sigma \rightarrow \mathbb{R}^+$.
$w_{del}(a)$of $a$ can be denoted as $w(a \rightarrow \lambda)$.
\item\label{def:string-cf-substitution} The cost function for substitution is $w_{sub}: \Sigma^2 \rightarrow \mathbb{R}_{\ge 0}$.
$w_{sub}(a, b)$ from $a$ to $b$ can be denoted as $w(a \rightarrow b)$.
\end{enumerate}
Thus, $w(\cdot \rightarrow \cdot)$ represents a notation that integrates the cost functions for three operations.
The cost function has the following properties:
\begin{enumerate}
\item\label{def:string-cf-positive} For all $a \in \Sigma$, $w(a \rightarrow a) = 0$. In all other cases, the weights are positive.
\item\label{def:string-cf-triangle} For all $a, b, c \in \Sigma \cup \{\lambda\}$, it holds that $w(a \rightarrow c) \leq w(a \rightarrow b) + w(b \rightarrow c)$.
In other words, the function satisfies the triangle inequality.
\end{enumerate}
\end{definition}

\begin{definition}[String edit distance]\label{def:string-edit-sequence}
A \emph{string edit sequence} from $x$ to $y$ is a sequence of
strings, where the first term is $x$, the last term is $y$ and
each term between $x$ and $y$ is attained by an edit operation from the previous term.
Naturally, the cost of a string edit sequence is
the sum of each edit operation cost applied in the edit sequence and
the \emph{edit distance} from $x$ to $y$ is the minimum cost of a string edit sequence from
$x$ to $y$.
\end{definition}

Alternatively, we align characters of two strings $x$ and $y$
so that each character of $x$ is matched to the counterpart of $y$.
Then, the cost of an alignment and edit distance can be naturally attained.
For instance, let $x=abbac$ and $y=abbda$.
The cost of an example alignment for the following
\begin{equation*}
\begin{array}{ccccccc}
\centering
     x = &a & b & b & a         & c & \lambda  \\
     y = &a & b & b & \lambda   & d & a
\end{array}
\end{equation*}
is $w(a\to \lambda)+w(c\to d)+w(\lambda\to a)$.
Now, the edit distance is the minimum cost of an alignment.

\begin{definition}[Alignment]\label{def:string-alignment}
An \emph{alignment} is a binary relation between positions in $x$ and $y$.
For a binary relation~$T\subseteq\{1,2,\cdots ,|x|\}\times \{1,2,\cdots ,|y|\}$ to be a valid alignment,
$T$ should satisfy the following:
for $\forall (i_1,j_1)\not=(i_2,j_2)\in T$, 1) $i_1\not= i_2, j_1\not = j_2$ and 2) $i_1<i_2$ implies $j_1<j_2$.
\end{definition}

\section{Exponent-String}\label{sec:exponent-string}
We propose \emph{\es{}} as a natural extension of strings to represent much wider information,
inspired by its run-length encoded format. 
Note that we assumed addition for the binary operation of 
semigroups $\mathbb{N}, \mathbb{Q}^+$ and $\mathbb{R}^+$ 
in Section~\ref{ssec:mathematical-background}.

\begin{definition}[$S$-\es{}]\label{def:exponent-string}
\textnormal{
    Given an alphabet $\Sigma$ and a semigroup $S=(\mathcal{S},\bino{})$, 
    we define an \emph{$S$-\es}~$p$ over~$\Sigma$ to be
    a finite sequence~$p = \left( (\sigma_1, s_1), (\sigma_2, s_2), \ldots ,(\sigma_n, s_n) \right)$
    where, for each term~$(\sigma_i, s_i)$, the base character~$\sigma_i$ is in $\Sigma$
    and the exponent~$s_i$ is in $\mathcal{S}$. 
    The empty sequence is always a valid $S$-\es{} denoted by $\lambda$. 
    For a nonempty sequence, two consecutive terms should have different 
    base characters---$\sigma_i\not=\sigma_{i+1}$ for $1\leq i \leq n-1$.
    We denote the set of $S$-exponent-strings as $\Sigma_{S}^*$.
}
\end{definition}

\begin{figure}[h!]
    \centering
    \includegraphics[width=.9\textwidth ]{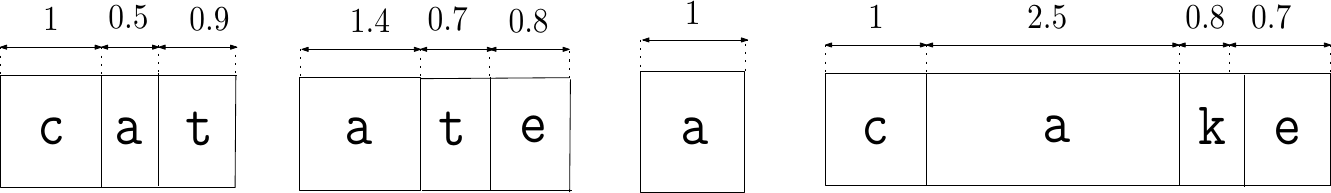}
    \caption{A pictorial illustration of four \reales{s}; $\mathtt{ca^{0.5}t^{0.9}}$, $\mathtt{a^{1.4}t^{0.7}e^{0.8}}$, $\mathtt{a^1}$ and $\mathtt{ca^{2.5}k^{0.8}e^{0.7}}$, 
    where the sound duration of each character is written as exponents. 
    Note that  $\mathtt{a}$'s in words \texttt{ate} and \texttt{cake} are long-vowels whereas 
    $a$ in \texttt{cat} is a 
    short-vowel.
    }
    \label{fig:exponent-example}
\end{figure}

\textbf{Representation.}
For an $S$-exponent-string~$p=\left((\sigma_1, s_1), (\sigma_2, s_2), \ldots, (\sigma_n, s_n)\right)$,
we can represent $p$ in the form~$p=a_1^{q_1}a_2^{q_2}\cdots a_k^{q_k}$
if and only if integers~$1=k_1< k_2< \cdots< k_{n+1}=k+1$ exist such that for each~$1\le i \le n$,
$s_i=q_{k_{i}}\bino{}q_{k_{i}+1}\bino{}\cdots\bino{}q_{k_{i+1}-1}$ and
$\sigma_i=a_{k_{i}}=a_{k_{i}+1}=\cdots=a_{k_{i+1}-1}$.
This setup allows $p$ to have multiple string form representations.
For instance, 
let $\mathcal{S}=\mathbb{R}^+$ and $\bino{}=+$. 
An $S$-\es{}~$p=\left((c,\frac{17}{7}),(a,\pi ),(b,3\sqrt{5})\right)$ can appear in different forms such as 
$p = c^1c^{\frac{10}{7}}a^{\pi}b^{3\sqrt{5}}$ or $p = c^{1.5} c^{0.5} c^{\frac{3}{7}}a^{\pi-2}a^{2}b^{\sqrt{5}}b^{2\sqrt{5}-0.5}b^{0.5}=\cdots$. 
Among various string representations of
$S$-\es{}~$p=\left((\sigma_1, s_1), (\sigma_2, s_2), \ldots ,(\sigma_n, s_n)\right)$,
we say the form~$p=\sigma_1^{s_1}\sigma_2^{s_2}\cdots \sigma_n^{s_n}$ 
without term-breaking in $p$
is a \emph{contraction form} of $p$.
It is important to recognize that expressions such as $p=c^1c^{\frac{10}{7}}a^{\pi}b^{3\sqrt{5}}$ loosely use notation
because $p$ is an $S$-\es{} and the right side is a particular representation of the exponent-string.
However, we demonstrate 
that such representation is not ambiguous; that is,
1. no two $S$-\es{} can have the same representation, 
and 2. every representation is associated with an $S$-\es{},
in Proposition~\ref{prop:representation-appropriate}.
\begin{lemma}\label{lem:unique-parsing}
    Given alphabet $\Sigma$,
    let $k\in\mathbb{N}$, $(a_1, a_2, \cdots, a_k)\in\Sigma^k$.
    Then, there exists a unique choice of integers $n, k_1, k_2,\cdots ,k_{n+1}$
    such that $k_1=1$, $k_{n+1}=k+1$, $k_i<k_{i+1}$,
    $a_{k_i}=a_{k_i+1}=\cdots=a_{k_{i+1}-1}$ 
    and $a_{k_j}\neq{}a_{k_{j+1}}$ 
    for $1\leq i \leq n$, $1\leq j \leq n-1 $.
\end{lemma}
\begin{proof}
    Let $X\coloneq{}\{i\mid{}a_{i-1}\neq{}a_{i} \textnormal{ for } 2\leq i\leq k\}$.
    Then, it is straightforward that $n=|X|+1$ and $k_2, k_3,\cdots, k_n$ are
    elements of $X$ in increasing order.
    As $X$ is well-defined for given $k$ and $a_1,a_2,\cdots,a_k$,
    each integer $n, k_1, k_2,\cdots ,k_{n+1}$ uniquely exists.
\end{proof}
\begin{proposition}\label{prop:representation-appropriate}
    Given alphabet $\Sigma$ and semigroup $S=(\mathcal{S},\bino{})$,
    let $k\in\mathbb{N}$, $(a_1, a_2, \cdots, a_k)\in\Sigma^k$, 
    and $(q_1,q_2,\cdots, q_k)\in\mathcal{S}^k$.
    Then, there exists a unique $S$-\es{} corresponding to 
    the representation $a_1^{q_1}a_2^{q_2}\cdots{}a_k^{q_k}$.
    In other words,
    there exist unique integer $n$ and sequence $\left((\sigma_1,s_1),(\sigma_2,s_2),
    \cdots,(\sigma_n,s_n)\right)\in\Sigma_S^*$
    satisfying:
    \begin{enumerate}
        \item\label{condition:prop-representation1} $\sigma_i\neq{}\sigma_{i+1}$ for $1\leq i \leq n-1$.
        \item\label{condition:prop-representation2} For some integers $k_1, k_2, \cdots, k_{n+1}$,
        \begin{enumerate}
            \item\label{condition:prop-representation2-a} $k_1=1$ and $k_{n+1}=k+1$,
            \item\label{condition:prop-representation2-b} $k_i<k_{i+1}$ for $1\leq i \leq n$,
            \item\label{condition:prop-representation2-c} $\sigma_i=a_{k_i}=a_{k_i+1}=\cdots =a_{k_{i+1}-1}$
            for $1\leq i \leq n$,
            \item\label{condition:prop-representation2-d} $s_i=q_{k_i}\bino{}q_{k_i+1}\bino{}\cdots \bino{}q_{k_{i+1}-1}$
            for $1\leq i \leq n$.
        \end{enumerate}
    \end{enumerate}
\end{proposition}
\begin{proof}
    From Lemma~\ref{lem:unique-parsing}, there exists a unique
    choice of $n, k_1, k_2,\cdots,k_{n+1}$
    that satisfies Conditions~\ref{condition:prop-representation1}, \ref{condition:prop-representation2-a},
    \ref{condition:prop-representation2-b} and \ref{condition:prop-representation2-c}.
    This directly characterizes $n$ and $\sigma_i$'s.
    Also, corresponding $s_i$'s from Condition~\ref{condition:prop-representation2-d}
    are uniquely determined as $s_i=q_{k_i}\bino{}q_{k_i+1}\bino{}\cdots \bino{}q_{k_{i+1}-1}$.
\end{proof}

\textbf{Factor.}
In the representation of a nonempty $S$-\es{}~$p=a_1^{s_1}a_2^{s_2}\cdots a_n^{s_n}$,
we consider each $a_i^{s_i}$ for $1\le i\le n$ as a \emph{factor} of $p$.
When base characters~$a_{i_1},a_{i_2}, \cdots, a_{i_k}$ are the
same---$\sigma=a_{i_1}=a_{i_2}= \cdots =a_{i_k}$---each corresponding
$a_{i_j}^{s_{i_j}}$ for $1\le j\le k$ is a \emph{$\sigma$-factor} of $p$.
For example, in a \rates{} representation~$p = a^{2.8}a^{1.3}b^2$, there are two types of factors:$a$-factor and $b$-factor,
with both $a^{2.8}$ and $a^{1.3}$ being $a$-factors.

\textbf{Concatenation.}
For two $S$-\es{s} $p=\left((\alpha_1,s_1),(\alpha_2,s_2),\cdots ,(\alpha_n,s_n)\right)$ 
and $q=((\beta_1,t_1),\allowbreak (\beta_2,t_2),\cdots ,(\beta_m,t_m))$,
concatenation $\cdot$ is defined as follows:
\begin{align*}
    &p\cdot{}q=\begin{cases}
        \left((\alpha_1,s_1),(\alpha_2,s_2),\cdots ,(\alpha_n,s_n\bino{}t_1),(\beta_2,t_2),\cdots ,(\beta_m,t_m)\right) 
        & \textnormal{ if } \alpha_n=\beta_1,\\
        \left((\alpha_1,s_1),(\alpha_2,s_2),\cdots ,(\alpha_n,s_n),(\beta_1, t_1),(\beta_2,t_2),\cdots ,(\beta_m,t_m)\right) 
        & \textnormal{ Otherwise.} \\
    \end{cases}\\
    &p\cdot \lambda = \lambda \cdot p = p.
\end{align*}
Concatenation can be described more intuitively in its representation 
as shown in Proposition~\ref{prop:representation-concat}.
\begin{proposition}\label{prop:representation-concat}
    For $S$-exponent-strings 
    $p=a_1^{x_1}a_2^{x_2}\cdots a_k^{x_k}\in\Sigma_S^*$ and
    $q=b_1^{y_1}b_2^{y_2}\cdots b_l^{y_l}\in\Sigma_S^*$, 
    $p\cdot q = a_1^{x_1}a_2^{x_2}\cdots a_k^{x_k}b_1^{y_1}\allowbreak b_2^{y_2}\cdots b_l^{y_l}$.
\end{proposition}
\begin{proof}
    Let $p=\left((\alpha_1,s_1),(\alpha_2,s_2),\cdots, (\alpha_n,s_n)\right)$ 
    and $q=\left((\beta_1,t_1),(\beta_2,t_2),\cdots ,(\beta_m,t_m)\right)$ in
    their formal sequence notation.
    If $\alpha_n\neq{}\beta_1$, 
    $p\cdot q = 
    \left((\alpha_1,s_1),(\alpha_2,s_2), 
    \cdots ,(\alpha_n,s_n),(\beta_1, t_1),(\beta_2,t_2),\cdots ,(\beta_m,t_m)\right)$.
    It is straightforward to see each exponent $s_i$ and $t_i$ can be split back to their
    corresponding $x_i$'s and $y_i$'s. 
    Thus,
    $a_1^{x_1}a_2^{x_2}\cdots a_k^{x_k}b_1^{y_1}b_2^{y_2}\cdots b_l^{y_l}$
    is appropriate representation.
    Otherwise, if $\alpha_n=\beta_1$, 
    $p\cdot q = 
    ((\alpha_1,s_1),(\alpha_2,s_2),\allowbreak
    \cdots ,(\alpha_n,s_n\bino{} t_1),(\beta_2,t_2),\cdots ,(\beta_m,t_m))$. 
    As $\bino{}$ is associative, every exponent including $s_n\bino{}t_1$ can be split
    back to its corresponding $x_i$'s and $y_i$'s.
    Therefore, $a_1^{x_1}a_2^{x_2}\cdots a_k^{x_k}b_1^{y_1}b_2^{y_2}\cdots b_l^{y_l}$
    is appropriate representation of $p\cdot q$.
\end{proof}
For example, if $S=(\mathbb{R},\times)$, $a^3b^{-\pi}\cdot b^2c^{4.1}
=a^3b^{-\pi}b^2c^{4/1}=a^3b^{-2\pi}c^{4.1}$.
For \rates{s} where the binary operation of $\mathbb{Q}^+$ is an addition, $c^{3.6}b^{2.4}b^{1}\cdot{}b^{0.2}a^{1.5}
=c^{3.6}b^{2.4}b^{1}b^{0.2}a^{1.5}=c^{3.6}b^{3.6}a^{1.5}$.

In practice, we often omit the concatenation symbol~$\cdot$ when its use is clear from the context and
we may also omit exponents of value 1 for simplicity.

As an extension of strings, we present Proposition~\ref{prop:basic-property} for \es{s}.
\begin{proposition}\label{prop:basic-property}
    Given an alphabet~$\Sigma$ and semigroup $S=(\mathcal{S},\bino{})$,
    \begin{enumerate}
        \item\label{prop:basic-property-1} $(\Sigma_S^*$,$\cdot)$ is a monoid where $\lambda$ is
        an identity element.
        \item\label{prop:basic-property-2} 
        If $S'$ is a subsemigroup of $S$, then 
        $(\Sigma_{S'}^*,\cdot_{S'})$ is a submonoid of $(\Sigma_{S}^*,\cdot_{S})$ where 
        $\cdot_S$ (respectively, $\cdot_{S'}$) is a concatenation 
        defined over $\Sigma_{S}^*$ (respectively, $\Sigma_{S'}^*$).
    \end{enumerate}
\end{proposition}
\begin{proof}
    (\ref{prop:basic-property-1}) By the definition of concatenation, $p\cdot \lambda =\lambda\cdot p= p$ for any $p\in \Sigma_S^*$.
    
    In order to show associativity, assume $p_1=a_1^{m_1}a_2^{m_2}\cdots a_t^{m_t},p_2=b_1^{n_1} 
    b_2^{n_2}\cdots b_k^{n_k}$ and
    $ p_3=c_1^{h_1}c_2^{h_2}\cdots c_l^{h_l}$. 
    From Proposition~\ref{prop:representation-concat}, 
    $(p_1\cdot p_2)\cdot{} p_3=p_1\cdot (p_2\cdot{} p_3)=a_1^{m_1}
    a_2^{m_2}\cdots a_t^{m_t}b_1^{n_1} 
    b_2^{n_2}\cdots b_k^{n_k}c_1^{h_1}c_2^{h_2}\cdots c_l^{h_l}$.
    If $p_1=\lambda$, $(p_1\cdot p_2)\cdot{} p_3=p_2\cdot p_3 = 
    p_1\cdot (p_2\cdot{} p_3)$. Similar can be shown for $p_2=\lambda$ 
    and $p_3=\lambda$.
    \\
    (\ref{prop:basic-property-2}) Let $S'=(\mathcal{S}',\bino{}')$.
    As $\mathcal{S'}\subseteq \mathcal{S}$, 
    we have $\Sigma_{S'}^*\subseteq \Sigma_{S}^*$. 
    Also, as $a\bino{}'b=a\bino{}b$ for $a,b\in\mathcal{S}'$, we have 
    $p\cdot_{S'} q =p\cdot_{S} q$ for $p,q\in\Sigma_{S'}^*$ from the
    definition of the concatenation.
\end{proof}
From the second property from Proposition~\ref{prop:basic-property},
$\Sigma_\mathbb{N}^*$ is a submonoid of \Qesset{}
and \Qesset{} is a submonoid of \Resset{}.
As $\Sigma_\mathbb{N}^*$ can be viewed as a set of strings in RLE format,
\Qesset{} and \Resset{} are reasonable extensions of strings.
We justify this formally in Proposition~\ref{prop:N-exp-string-iso},
that shows monoid $(\Sigma_\mathbb{N}^*,\cdot)$ is 
isomorphic with monoid of $\Sigma^*$.
\begin{proposition}\label{prop:N-exp-string-iso}
    The monoid $(\Sigma_\mathbb{N}^*,\cdot)$ is isomorphic with 
    $\Sigma^*$ equipped with string concatenation
    \footnote{As concatenation was formally defined for 
    exponent-strings, 
    we use the term string concatenation to refer to the
    concatenation of traditional strings.}.
\end{proposition}
\begin{proof}
    Consider the following mapping 
    $F:\Sigma_{\mathbb{N}}^*\rightarrow\Sigma^*$ defined as follows:
    \begin{align*}
        &F(\lambda)=\lambda,\\
        &F(a_1^{n_1}a_2^{n_2}\cdots a_k^{n_k})=\overbrace{a_1\cdots a_1}^{n_1}\overbrace{a_2\cdots a_2}^{n_2}\cdots \overbrace{a_k \cdots a_k}^{n_k}.
    \end{align*}
    Then, $F$ is bijective and $F(p\cdot q)= F(p)F(q)$. 
    Therefore, $F$ is an isomorphism from $\Sigma_{\mathbb{N}}^*$ to $\Sigma^*$.
\end{proof}
\begin{remark}
In the later sections, we use \emph{$\mathbb{N}$-\es{}} and \emph{string} interchangeably.
Consequently, we may apply well-known properties of $\Sigma^*$ to $\Sigma_{\mathbb{N}}^*$.
\end{remark}

In the later parts of the paper, our discussion focuses mostly on \reales{s} and \rates{s}, 
starting by giving a natural extension on string terminologies for \reales{}.
Note that they are also applicable to \rates{}.

\textbf{Length, Factor-Length of Contraction Form.} 
For $p=\sigma_1^{s_1}\sigma_2^{s_2}\cdots\sigma_n^{s_n}\in$\Resset{}, 
the length~$len(p)$ of $p$ is defined as $len(p):=\sum_{i=1}^n s_i$.
The factor length~$Flen(p)$ of $p$ is the number of factors in the contraction form of $p$.
Specifically, if a \reales{}~$p=\tau_1^{t_1}\tau_2^{t_2}\cdots\tau_k^{t_k}$ is given in the contraction form,
then $Flen(p)=k$.

It follows directly that the length of an $\mathbb{N}$-\es{} matches the length of a traditional string,
and Flen of an $\mathbb{N}$-\es{} corresponds to the number of runs in RLE-strings.

\textbf{Bracket Notation.} 
For $p=\sigma_1^{s_1}\sigma_2^{s_2}\cdots\sigma_n^{s_n}$, 
a function $f_p: \{x \mid 0 \leq x < len(p)\} \rightarrow \Sigma $ is a base character
of position $x$.
\begin{equation*}
f_p(x) = \left\{
\begin{array}{rl}
    \sigma_1, & \text{if } 0\leq x < s_1,\\
    \sigma_2, & \text{if } s_1 \leq x < s_1+s_2,\\
    &\vdots \\
    \sigma_n, & \text{if } \displaystyle{\sum_{i=1}^{n-1}s_i \leq x < len(p)=\sum_{i=1}^{n}s_i}.
    \end{array} \right.
\end{equation*}
In the spirit of extending the bracket notation, we denote $p[x]$ instead of $f_p(x)$.

The choice to consider only non-negative real numbers for the length function is because it simplifies the identification of character positions.
For instance, considering a hypothetical sequence $p = a^2b^{-2.5}$,
our terminology eliminates the need to determine whether 
$p[1.2]=a$ or $p[1.2]=p\left[2+(-0.8)\right]=b$.
This approach also prevents complications associated with potential negative lengths or fractional indices,
thereby ensuring all character positions are non-negative and bounded by the defined string length.

Furthermore, in the context of $p[x]$, where $p[0]$ is defined but $p[\text{len}(p)]$ is not,
this follows traditional string indexing conventions.
The end index is exclusive, aligning with typical string handling where accessing the position at the length of the string is out of bounds.
This convention streamlines the implementation of the function by removing the need to handle edge cases for indices outside the valid range.

\textbf{Prefix, Infix and Suffix.} 
For $p,q\in$ \Resset{},
$q$ is an infix (prefix, suffix) of $p$ if 
there exist $u,v\in$ \Resset{} such that $p=uqv$ ($p=qv$, $p=uq$, respectively).
For instance,
a \reales{}~$\mathtt{a^{2.3}uto^{\frac{12}{7}}ma^{1.1}ta^{1.3}}$
has a prefix~$\mathtt{a^{2.3}uto}$,
infix~$\mathtt{o^{\frac{5}{7}}ma}$
and suffix~$\mathtt{a^{0.1}ta^{1.3}}$.
    
\section{Edit Distance of Exponent-Strings}\label{sec:dp-real-exponent-string}
Now, having established the foundational concepts on \es{s},
we are prepared to explore edit distance for two \reales{s}. 

\subsection{Fundamentals of Exp-Edit Distance}\label{ssec:expdist-seq}

In analogy with the traditional string edit distance,
let us define a set of production rules for transforming one \reales{} to another
by replacing an infix~$x$ to another infix~$y$.
We categorize the operations permissible within these rules, known as \emph{exp-edit operations},
into three types as follows:
\begin{enumerate}
\item Insert operation: $\lambda \rightarrow a^q$ for arbitrary $a\in \Sigma$ and $q\in \mathbb{R}^{+}$.
\item Delete operation: $a^q \rightarrow \lambda$ for arbitrary $a\in \Sigma$ and $q\in \mathbb{R}^{+}$.
\item Substitute operation: $a^q \rightarrow b^q$ for arbitrary $a, b\in \Sigma$ and $q\in \mathbb{R}^{+}$.
\end{enumerate}

We denote $p \Rightarrow q$ for applying edit operations to $p$ and obtain $q$ as the following equation:
\begin{equation*}
    p\Rightarrow q \textnormal{ if and only if } p=p_1xp_2, q=p_1yp_2 
    \textnormal{ and } x\rightarrow y 
\end{equation*}
We say that $p$ has resulted in $q$ via the $x\rightarrow y$ exp-edit operation,
in the spirit of accepting exp-edit operations as a procedure of editing one \reales{} to another.
Given two \reales{s}~$u$ and $v$,
if a finite sequence of \reales{s}~$\{r_i\}_{0\leq i \leq n}$ satisfies 
$u=r_0$, $v=r_n$ and $r_{k}\Rightarrow r_{k+1}$ for every $0\leq k <n\in \mathbb{N}$, then
we call $\{r_i\}_{0\leq i \leq n}$ an \emph{exp-edit sequence} from $u$ to $v$.

We extend the cost of string edit operations from Definition~\ref{def:string-cost-function}
to define the costs for exp-edit operations and sequences.
Assume cost functions $w_{ins}: \Sigma \rightarrow \mathbb{R}^{+}, w_{del}: \Sigma \rightarrow 
\mathbb{R}^{+}, w_{sub}:\Sigma^2 \rightarrow \mathbb{R}_{\geq 0}$
for arbitrary $a, b\in \Sigma$ and $q\in \mathbb{R}^{+}$
are given as follows:
\begin{itemize}
   \item $w(\lambda \rightarrow b^q)= q\times w(\lambda \rightarrow b)$,
   \item $w(a^q \rightarrow \lambda)= q\times w(a \rightarrow \lambda)$,
   \item $w(a^q \rightarrow b^q)= q\times w(a \rightarrow b)$.
\end{itemize}

For an arbitrary exp-edit sequence~$K=\{r_i\}_{0\leq i \leq n}$, 
we define cost~$W(K)$ of the sequence~$K$ or $W(r_0\Rightarrow r_1\Rightarrow \cdots\Rightarrow r_n)$
inductively as follows:
\begin{itemize}
   \item $W(r_0\Rightarrow r_1)=w(x\rightarrow y)$ if $r_0 \Rightarrow r_1$ via $x\rightarrow y$
   \item For $k\in \mathbb{N}$, 
   $W(r_0\Rightarrow \cdots\Rightarrow r_k\Rightarrow r_{k+1})=
   W(r_0\Rightarrow \cdots\Rightarrow r_k)+w(x\rightarrow y)$
   if  $r_k \Rightarrow r_{k+1}$ via $x\rightarrow y$
\end{itemize}

\begin{definition}\label{def:pseudo-edit-dist}
\textnormal{
    For \reales{s}~$u,v\in \Sigma_{\mathbb{R}^{+}}^*$, let $\mathbb{SEQ}_{u,v}$ be the set
    of every exp-edit sequence from $u$ to $v$. 
    We define exp-edit distance or $dist(u,v)$ to be $\inf \left(W(\mathbb{SEQ}_{u,v})\right)$.
    }
\end{definition}

As $W(\mathbb{SEQ}_{u,v})$ is neither finite nor some set of integers with a lower bound,
it is not trivial to assert the existence of an exp-edit sequence from $u$ to $v$ with a cost~$dist(u,v)$;
that is, we cannot readily assume that $dist(u,v)$ is the minimum cost, unlike for the 
case of strings as the number of string alignments is finite.
Definition~\ref{def:pseudo-edit-dist} acknowledges such issue by choosing $dist(u,v):=\inf \left(W(\mathbb{SEQ}_{u,v})\right)$ instead of $\min \left(W(\mathbb{SEQ}_{u,v})\right)$. Also, $dist(u,v)$ is always guaranteed to exist and is unique.
It is because the cost of any exp-edit sequence has an obvious lower bound~0 and
due to the Proposition~\ref{prop:glb-property}.
We will revisit this topic in Section~\ref{sec:algo-for-exp-edit-dist}. 

\subsection{Extension of String Alignment Through Exp-Matching}\label{ssec:exp-matching}
We can think of an alternative definition of exp-edit distance, similar to the string alignment,
that is equivalent to Definition~\ref{def:pseudo-edit-dist}.
We adapt string alignment concepts to a continuous framework to extend it to exp-edit distance in Definition~\ref{def:alter-pseudo-edit-dist}.

In string alignment, represented as $T\subseteq \{1,2, ...n \}\times \{1,2, ...m\}$ where $n$ and $m$ are length of strings,
alignment conditions ensure that matches reflect the actual edit sequences of the string.
Similarly, in exp-matching, represented as $T\subseteq [0,L_1] \times [0,L_2]$,
Conditions~3~and~4 directly correspond to the conditions in Definition~\ref{def:string-alignment}.
Condition~1 simplifies the class of relations we need to consider,
particularly to ensure that we can compute the cost of every path and
to eliminate any potential for impossible or impractical alignments.
Condition~2 addresses cases where matching can \emph{stretch} a small amount of symbol.
For instance, consider when transforming from $a^1$ to $a^{100}$,
and matching is $\{(x,y) \mid y=100x, x\in [0,1]\}$, stretching 1 unit of $a$ to 100 units of $a$.
Then the cost would be 0 since every position would match the same symbol.

Thus, exp-matching requires additional conditions, extending from string alignment, due to its continuous nature.
This transition also shifts the cost calculation to an integral.

\begin{definition}\label{def:alter-pseudo-edit-dist}
\textnormal{
    Assume alphabet $\Sigma$ and cost functions $w_{del}:\Sigma\rightarrow\mathbb{R}^{+}, w_{ins}:\Sigma\rightarrow\mathbb{R}^{+}, w_{sub}:\Sigma\times\Sigma\rightarrow\mathbb{R}_{\geq 0}$ are given.
    Consider two \reales{s}~$p_1,p_2$ over $\Sigma$ and $L_1=len(p_1), L_2=len(p_2)$. 
    Then, $E\subseteq [0,L_1)\times [0,L_2)$ is an exp-matching from $p_1$ to $p_2$ if it satisfies following:
    \begin{enumerate}
        \item $E$ is a union of finitely many line segments~(Assume terminology of
        Euclidean space as $[0,L_1)\times [0,L_2)$ is a rectangle in two-dimensional Euclidean space).
        \item Each line segment is parallel to the identity line(y=x line).
        \item\label{item:no-overlapping} Let $E=E_1\cup E_2 \cup \cdots \cup E_n$ where each $E_i$ are line segments.
        Let $X_{1}, X_{2}, \cdots , X_{n}\subseteq [0,L_1)$ be projection of each line segments to X-axis and
        let $Y_{1}, Y_{2}, \cdots , Y_{n}\subseteq [0,L_2)$ be projection to Y-axis.
        Then $X_{1}, X_{2}, \cdots , X_{n}$ are disjoint and $Y_{1}, Y_{2}, \cdots , Y_{n}$ are also disjoint.
        \item\label{item:no-crossing} There does not exist element $(x_1,y_1),(x_2,y_2)\in E$ such that $x_1<x_2$ and $y_1>y_2$.
        In other words, $x_1<x_2$ implies $y_1\leq y_2$.
    \end{enumerate}
    Let $E_X$ be the projection of $E$ to the X-axis and $E_Y$ be the projection to the Y-axis. The cost of such exp-matching~$\overline{W}(E)$ is defined as sum of three values:
    \begin{enumerate}
        \item $\int_{[0,L_1)-E_X}w_{del}(p_1[x])dx$
        \item $\int_{[0,L_2)-E_Y}w_{ins}(p_2[y])dy$
        \item $\frac{1}{\sqrt{2}}\int_{E} w_{sub}(p_1[r_x],p_2[r_y])d\vec{r}$ 
        where $\vec{r}=(r_x,r_y)$.
    \end{enumerate}
    Exp-edit distance from $p_1$ to $p_2$ is an infimum possible cost of exp-matching.
}
\end{definition}

String alignment from Definition~\ref{def:string-alignment} is a binary relation between a set of integers so that 
for an alignment~$T$ between strings $w$ and $s$, $(i,j)\in T$ implies
that characters $w[i-1]$ and $s[j-1]$ were matched.
Similarly, for exp-matching $E$ from $p_1$ to $p_2$,
$(x,y)\in E$ implies that the position~$x$ in $p_1$ is matched to the position~$y$ in $p_2$.
Likewise, the unmatched positions($[0,L_1)-E_X$ and $[0,L_2)-E_Y$) should be handled by deletion and insertion.
Therefore, $\int_{[0,L_1)-E_X}w_{del}(p_1[x])dx$, $\int_{[0,L_2)-E_Y}w_{ins}(p_2[y])dy$ and 
$\frac{1}{\sqrt{2}}\int_{E} w_{sub}(p_1[r_x],p_2[r_y])d\vec{r}$
are the cost of deletion, insertion and substitution respectively.
\begin{figure}[h!]
    \centering
    \includegraphics[width=1\textwidth ]{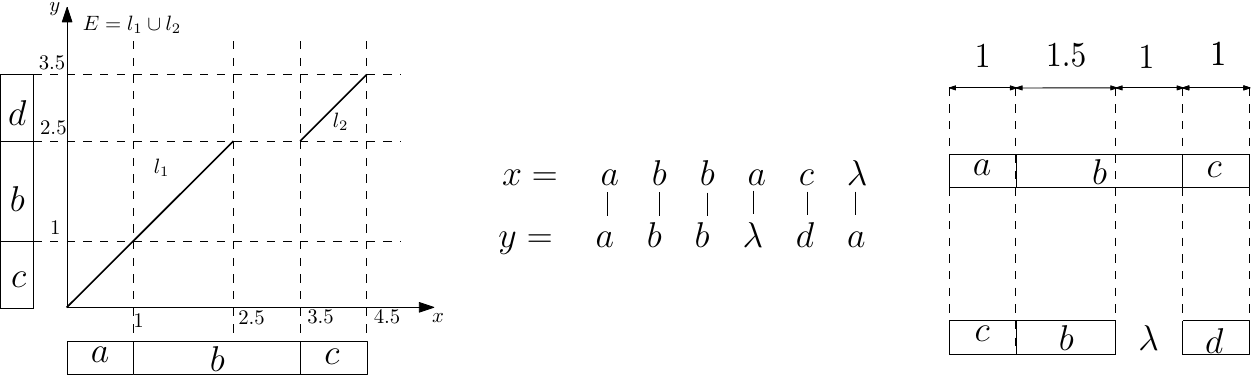}
    \caption{
    An illustration that compares an exp-matching and string alignment.
    From the left, each picture depicts
    an exp-matching~$E$ from $ab^{2.5}c$ to $cb^{1.5}d$,
    an alignment from $abbac$ to $abbda$ and finally,
    an alignment of $E$.
    }
    \label{fig:example-alignment-trace}
\end{figure}
In order to prove Definitions~\ref{def:pseudo-edit-dist} and \ref{def:alter-pseudo-edit-dist} are 
equivalent, Theorems~\ref{thm:alter-def-equiv-upper} and \ref{thm:alter-def-equiv-lower} are given in the following section.
\begin{theorem}\label{thm:alter-def-equiv-upper}
    Let $u,v\in \Sigma_{\mathbb{R}^{+}}^*$ be \reales{s}. 
    For an arbitrary exp-edit sequence $K$ from $u$ to $v$,
    there exists an exp-matching $E$ such that $\overline{W}(E)\leq W(K)$.
\end{theorem}
\begin{proof}
Consider an arbitrary exp-edit sequence as a transformation that maps positions from the
source~\reales{}~$u$ to the destination~\reales{}~$v$.
A delete operation~$p_1a^qp_2 \rightarrow p_1p_2$ corresponds to the transformation~$T$ that works as follows:
\begin{enumerate}
\item $T(x) = x$ for $0\le x < len(p_1)$,
\item $T(x)$ is undefined for $ len(p_1) \le x < len(p_1)+q$,
\item $T(x) = x-q$ for $len(p_1)+q \le x < len(p_1)+q+len(p_2)$.
\end{enumerate}
A substitute operation~$p_1a^qp_2 \rightarrow p_1b^qp_2$ corresponds to $T(x)=x$ for $0\le x < len(p_1)+q+len(p_2)$,
and an insert operation~$p_1p_2 \rightarrow p_1a^qp_2$ correspond to 
\begin{enumerate}
\item $T(x) = x$ for $0\le x < len(p_1)$,
\item $T(x) = x+q$ for $len(p_1) \le x < len(p_1)+len(p_2)$.
\end{enumerate}

Let $\{K_i\}_{1\le i \le n}$ be an exp-edit sequence and let $T_i$ be the transformation corresponding to $K_i \Rightarrow K_{i+1}$.
Define $T := T_1 \circ T_2 \cdots \circ T_{n-1}$
and consider the set~$E := \{(x,T(x))\mid x\in domain(T)\}$, which can be recognized as an exp-matching.

By the triangle inequality,
the exp-edit sequence of directly substituting positions from $u$ that are in $domain(T)$ to their corresponding positions in $v$
costs no more than the exp-edit sequence that involves multiple substitutions.
Similarly, for positions from $[0,len(u)) \setminus domain(T)$~($[0,len(v)) \setminus range(T)$),
a direct insertion (deletion, respectively) costs less than or equal to the exp-edit sequence that involves multiple substitutions.

Thus, the cost of $E$ is less than or equal to that of $\{K_i\}_{1\leq i \leq n}$.
\begin{figure}[htb!]
    \centering
    \includegraphics[width=0.4\linewidth]{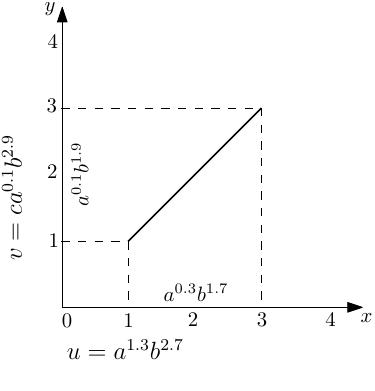}
    \caption{$a^{0.1}b^{1.9}$ and $a^{0.3}b^{1.7}$ are infixes corresponding to the line segment in the figure.}
    \label{fig:alter-def-upper}
\end{figure}
\end{proof}

\begin{theorem}\label{thm:alter-def-equiv-lower}
    Let $u,v\in \Sigma_{\mathbb{R}^{+}}^*$ be \reales{s}. 
    For an arbitrary exp-matching $E$ of $u$ and $v$, there exists an exp-edit sequence $K$ from $u$ to $v$ such that $\overline{W}(E) = W(K)$.
\end{theorem}
\begin{proof}
   Given an exp-matching~$E$, one can construct an exp-edit sequence that
    (1) substitute every factor of positions in $E_X$ to the factor of corresponding positions in $E_Y$,
    (2) delete every factor in positions~$[0,len(u))-E_X$ and
    (3) insert every factor in positions~$[0,len(v))-E_X$. 
    The cost of such an exp-edit sequence is exactly the cost of $E$.
\end{proof}
\section{Algorithm for Exp-Edit Distance}\label{sec:algo-for-exp-edit-dist}
Concerning our motivation to study exp-edit distance, that was
to enhance automatic diagnosis of paraphasia,
coming up with an algorithm is crucial.
We propose an algorithm for computing exp-edit distance,
for the case of \rates{s}.
\begin{remark}
    Although formal definitions and theorems about exp-edit distance
    work on the domain of \reales{}, we consider only \rates{s} as an instance
    of the exp-edit distance problem. 
    This is because (1) real numbers cannot be precisely represented 
    in computation models such as Turing machine, (2) approximate representation with arbitrary precision is possible by rational 
    numbers and (3) the algorithm is based on the property of rational number.
\end{remark}

\RestyleAlgo{ruled}
\begin{algorithm}[H]
\caption{Computation of exp-edit distance}
\label{algo:exp-edit distance calculation}
\SetAlgoLined
\KwIn{\rates{s} $p=a_1^{x_1}a_2^{x_2}\cdots a_n^{x_n}$ and $q=b_1^{y_1}b_2^{y_2}\cdots b_m^{y_m}$}
\KwOut{Exp-edit distance from $p$ to $q$}

\tcp{Step 0: Addressing exceptional cases}
\BlankLine
\uIf{$p=\lambda$ and $q\neq{}\lambda$}{
    return $\sum_{i=1}^m y_i\times{}w_{ins}(b_i)$\;
}
\uElseIf{$p\neq{}\lambda$ and $q=\lambda$}{
    return $\sum_{i=1}^n x_i\times{}w_{del}(a_i)$\;
}
\ElseIf{$p=\lambda$ and $q=\lambda$}{
    return 0\;
}
\BlankLine
\tcp{Step 1: Convert $p$ and $q$ to RLE strings}
\BlankLine
Convert $p$ and $q$ into contraction form\;
\BlankLine
$C\leftarrow{}$ Common denominator of every exponent of $p$ and $q$\;
\BlankLine
$w_1\leftarrow{}$ RLE string attained by multiplying $C$
to each exponent of $p$\;
\BlankLine
$w_2\leftarrow{}$ RLE string attained by multiplying $C$
to each exponent of $q$\;
\BlankLine
\tcp{Step 2: Compute (RLE) string edit distance}
\BlankLine
$d\leftarrow{}$ String edit distance from $w_1$ to $w_2$\;
\BlankLine
\tcp{Step 3: Return exp-edit distance}
\BlankLine
return $d/C$\;
\BlankLine
\end{algorithm}
Although Algorithm~\ref{algo:exp-edit distance calculation}
is described in a straightforward way of converting \rates{s}
to string and use string edit distance to calculate exp-edit distance,
proof of its correctness requires specification about following points.:
\begin{enumerate}
    \item \textbf{Existence of minimum cost exp-matching:}
    Both Definitions~\ref{def:pseudo-edit-dist} and 
    \ref{def:alter-pseudo-edit-dist} are defined as an infimum cost.
    Initially addressed in Section~\ref{ssec:expdist-seq},
    it is not trivial whether there exists exp-edit sequence  
    from one \reales{} to another with minimum cost because 
    of the infinite number of possible exp-edit sequences.
    In order to demonstrate 
    non-triviality of minimum cost exp-edit sequence, consider 
    the following competition between Alice and Bob:
    
    Given $p,q\in$\Resset{} and $n\in\mathbb{N}$, 
    Alice initiates the game by coming up with an exp-edit sequence
    from $p$ to $q$, using $n$ many exp-edit operations.
    Then, Bob comes up 
    with an exp-edit sequence using $n+1$ many exp-edit operation.
    If Bob came up with a lower cost exp-edit sequence than Alice, 
    Bob wins. Otherwise, Alice wins.
    
    At first glance, above game seems to be advantageous
    for Bob, as Bob is allowed to have more refined
    exp-edit sequence.
    However, that is not the case if Alice can come up with 
    a minimum cost exp-edit sequence, meaning that, for large enough 
    $n$, Alice can always win.
    We address this issue in Theorem~\ref{thm:inf-min}, establishing
    that a minimum cost exp-edit sequence---a sequence with specified exp-edit distance---always exists.
    \item \textbf{Relation between string edit distance and 
    exp-edit distance:} 
    Step 2 of Algorithm~\ref{algo:exp-edit distance calculation}
    computes string edit distance to finally compute exp-edit distance
    at Step 3.
    Although Definition~\ref{def:pseudo-edit-dist} is 
    a natural way to extend string edit distance, it allows
    non-discrete edits such as deleting 0.1 units of character,
    while traditional string edit operations are restricted
    to apply edit operation to one character.
    Therefore, clearly identifying a relation between 
    string edit distance and exp-edit distance is crucial.
    Corollary~\ref{cor:string-pseudo-equiv} shows
    that exp-edit distance between two $\mathbb{N}$-exponent-strings
    is equal to string edit distance. 
    Besides providing formal justification for Algorithm~\ref{algo:exp-edit distance calculation},
    Corollary~\ref{cor:string-pseudo-equiv} has its 
    significance in showing exp-edit distance is an
    extension of string edit distance.
    
    \item \textbf{Identifying how exp-edit distance changes 
    when exponents of $p$ and $q$ are multiplied by $C$:}
    At Step 1 of Algorithm~\ref{algo:exp-edit distance calculation},
    it multiplies some integer $C$ on each exponent of $p$ and $q$, 
    resulting in $w_1$ and $w_2$.
    Theorem~\ref{thm:similar} guarantees that exp-edit distance 
    from $w_1$ to $w_2$ is $C$ times greater than exp-edit distance
    from $p$ and $q$.
\end{enumerate}
\subsection{Proof of Correctness}\label{ssec:proof-correctness}
We start from the first point which gives useful insights for the 
other two points.

\begin{definition}\label{def:box}
\textnormal{
    Given two \reales{s}~$p_1$ and $p_2$,
    let $p_1=a_1^{c_1}\cdots a_n^{c_n}$ and $p_2=b_1^{d_1}\cdots b_k^{d_k}$ be their contraction form
    where $n,k\in \mathbb{Z}$, $a_i,b_j\in \Sigma$ and 
    $c_i,d_j\in\mathbb{R}^{+}$ for every $1\leq i\leq n, 1\leq j\leq k $. 
    A \emph{box} of $p_1, p_2$ is a term to refer any rectangle in $[0,len(p_1)]\times [0,len(p_2)]$ formed by vertical lines 
    $x=\sum_{i=0}^t c_i$, $x=\sum_{i=0}^{t+1} c_i$ 
    and horizontal lines $y=\sum_{i=0}^{h} d_i$, $y=\sum_{i=0}^{h+1} d_i$
    for every $0\leq t< n $ and $0\leq h <k$, where $c_0=d_0=0$.
}
\end{definition}

The term `box' from Definition~\ref{def:box}
is an intuitive term for visualizing the area formed by \reales{s} $p_1$ and $p_2$.
We can see that arbitrary points $(x_1,y_1),(x_2,y_2)$ in interior of some single box of $p_1, p_2$
satisfies $p_1[x_1]=p_1[x_2], p_1[y_1]=p_1[y_2], p_2[x_1]=p_2[x_2]$ and $p_2[y_1]=p_2[y_2]$. 
\begin{theorem}\label{thm:inf-min}
    For two \reales{s}~$p_1, p_2$, there exists an exp-matching such that $\overline{W}(E)=dist(p_1,p_2)$.
\end{theorem}
\begin{proof}
\textbf{Step 1}: Basic notations of proof

    If one of $p_1$ and $p_2$ is $\lambda$, our theorem is trivially true as inserting or deleting every symbol attains minimum cost, by triangle inequality.
    Let $p_1=\sigma_1^{c_1}\cdots \sigma_n^{c_n}$ and $p_2=\gamma_1^{d_1}\cdots \gamma_k^{d_k}$ at its contraction form, and let $c_0=d_0=0$. 
    For $1\leq i \leq n$ and $1\leq j \leq k$, let $b_{i,j}$ denote the set of points 
    that is at the interior and boundary of the box that is formed by vertical lines 
    $x=\sum_{v=0}^{i-1} c_v$, $x=\sum_{v=0}^{i} c_v$ 
    and horizontal lines $y=\sum_{v=0}^{h-1} d_v$, $y=\sum_{v=0}^{h} d_v$.
    In Figures~\ref{fig:thm min inf alg} and \ref{fig:thm min inf}, horizontal and vertical lines that form the box are denoted as dashed lines.
    Also, we define corresponding characters of $b_{i,j}$ as $P_1(b_{i,j})=\sigma_i$ and $P_2(b_{i,j})=\gamma_j$. 
    For instance, if $p_1=a^{2.7}b^3$ and $p_2=c^{4.3}d^{2.4}$, 
    $P_1(b_{2,1})=b$ and $P_2(b_{2,1})=c$.
    Let $U(p_1,p_2)$ denote the set of every exp-matching from $p_1$ to $p_2$.
    For convenience, we denote $U(p_1,p_2)$ as $U$.
    Our strategy for this proof is to consider a subset of $U$, 
    and argue that a specific subset of $U$ contains an exp-matching
    of cost $dist(p_1,p_2)$. 
    
\textbf{Step 2}: Reducing to Subset of $U$-part 1

    As Definition~\ref{def:alter-pseudo-edit-dist} is infimum cost of exp-matching,
    we may consider the sequence of exp-matching $\{E_i\}_{i\in \mathbb{N}}$ 
    where its costs converge to $dist(p_1,p_2)$. 
    \begin{mdframed}
    \textbf{Lemma}
    There exists $B$,
    which is subset of $\{b_{i,j}|1\leq i \leq n,1\leq j \leq k \in\mathbb{N} \}$
    and satisfies 1) $\forall b_{i_0,j_0},b_{i_1,j_1}\in B,
    i_0<i_1 \implies j_0\leq j_1$ and
    2) subsequence of $\{E_i\}_{i\in \mathbb{N}}$
    where the line segments in each exp-matching are covered by $B$
    (in other words, is a subset of $\bigcup_{b\in B} b$), is 
    an infinite subsequence.\\
    \textbf{Proof}
    For $v\in\mathbb{N}$, let $B_v$ be the minimal subset of
    $\{b_{i,j}|1\leq i \leq n,1\leq j \leq k \in\mathbb{N} \}$
    where $E_v$'s line segments are a subset of $\bigcup_{b\in B_v} b$.
    In other words, $B_v$ is a minimal subset of
    $\{b_{i,j}|1\leq i \leq n,1\leq j \leq k \in\mathbb{N} \}$ 
    that covers every line segment of $E_v$. 
    Notice that if $\forall b_{i_0,j_0},b_{i_1,j_1}\in B_v, i_0<i_1 \implies j_0\leq j_1$ is violated, then Condition~4 of Definition~\ref{def:alter-pseudo-edit-dist} is also violated,
    because $B_v$ is a minimal subset so every box in $B_v$ covers some part of the line segment.
    Consider subsequences of $\{E_i\}_{i\in \mathbb{N}}$ divided by a criterion
    that whether each term has the same $B_v$.
    As there are only $nk$ many boxes, there are only finitely many 
    possible $B_v$. 
    Namely, $2^{nk}$ becomes laid-back upper bound, if one considers whether $b_{i,j}\in B$ or $b_{i,j}\not\in B$ for every $1\leq i \leq n,1\leq j \leq k \in\mathbb{N}$.
    However, $\{E_i\}_{i\in \mathbb{N}}$ is an infinite sequence.
    Therefore, by the pigeonhole principle, there exists some choice of $B_v$ such that the subsequence of $\{E_i\}_{i\in \mathbb{N}}$
    where the line segments in each exp-matching are covered by $B$ is 
    an infinite subsequence.\qed
    \end{mdframed}
    
    Let $\{E_i'\}_{i\in \mathbb{N}}$ be the infinite subsequence of $\{E_i\}_{i\in \mathbb{N}}$ 
    where each line segments are a subset of $\bigcup_{b\in B} b$.
    \begin{mdframed}
    \textbf{Definition} For \reales{s} $x,y$ and set $\mathcal{B}\subseteq 
    \{b_{i,j}|1\leq i \leq Flen(x),1\leq j \leq Flen(y) \in\mathbb{N} \}$, 
    $T_0(\mathcal{B},x,y)$ is the set of exp-matching in $U(x,y)$ where line segments are covered by $\mathcal{B}$.
    \end{mdframed}
    For convenience, we denote $T_0(B,p_1,p_2)$ to $T_0$.
    We have that $E_i'\in T_0$ for every $i\in\mathbb{N}$.
    Then, as $\lim_{i\to\infty}E_i'=\lim_{i\to\infty}E_i=dist(p_1,p_2)$, we have
    the that $dist(p_1,p_2)=\inf \overline{W}(U)=\inf \overline{W}(T_0)$.
    Therefore, proving that there exists 
    an exp-matching in $T_0$ that admits minimum cost among every exp-matching of $T_0$ proves our theorem.
    
\textbf{Step 3}: Reducing to Subset of $U$-part 2

    Now consider the following procedure that can be applied to any exp-matching.
    Assume some exp-matching $X$ is given.
    
    \RestyleAlgo{ruled}
    \begin{algorithm}[H]
    \caption{Normalization Procedure}
    \label{algo:segment_clipping}
    \SetAlgoLined
    \KwIn{An exp-matching~$X$ which is a set of line segments}
    \KwOut{A set of line segments contained within the bounding box}
    
    \BlankLine
    
    \BlankLine
    \tcp{Step 1: Clip segments crossing the box boundary}
    Cut every line segment in $X$ that crosses the boundary of the box by the intersection point with the boundary\;
    
    \BlankLine
    \tcp{Step 2: Translate segments horizontally}
    \For{each line segment $s$ from rightmost to leftmost}{
        Translate $s$ to the right until Condition~3 of Definition~\ref{def:alter-pseudo-edit-dist} is violated or $s$ starts to cross the box\;
    }
    
    \BlankLine
    \tcp{Step 3: Translate segments vertically}
    \For{each line segment $s$ from rightmost to leftmost}{
        Translate $s$ upwards until Condition~3 of Definition~\ref{def:alter-pseudo-edit-dist} is violated or $s$ starts to cross the box\;
    }
    
    \BlankLine
    \tcp{Step 4: Merge continuous segments}
    Merge continuous line segments within the boundary of the box\;
    \end{algorithm}
    
    Step by step example of this procedure is shown in Figure~\ref{fig:thm min inf alg}.
    \begin{figure}

    \includegraphics[width=0.5\textwidth ]{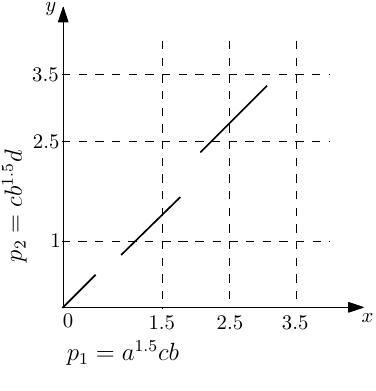}
        \centering
    \begin{tabular}{cc}
    \includegraphics[width=0.4\textwidth ]{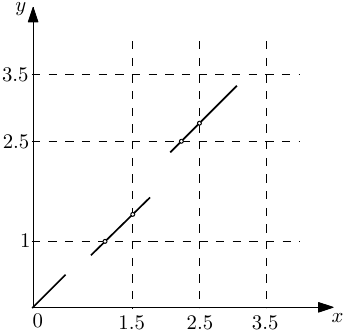}&
    \includegraphics[width=0.4\textwidth ]{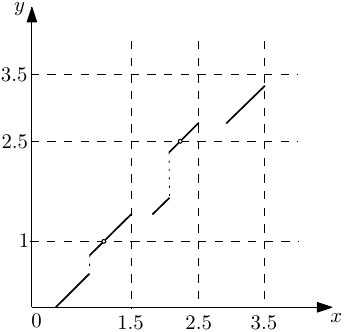}\\
    \includegraphics[width=0.4\textwidth ]{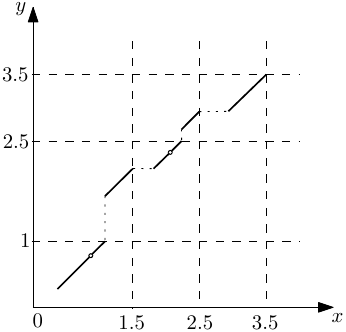}&
    \includegraphics[width=0.4\textwidth ]{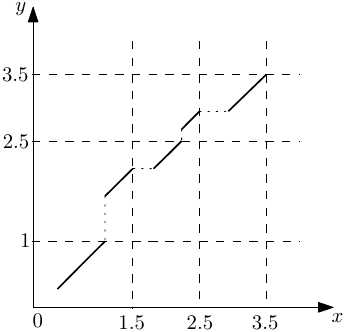}
    \end{tabular}
    \caption{
    Original exp-matching from $p_1=a^{1.5}cb$ to $p_2=cb^{1.5}d$ is shown at the top. 
    From left-upside to right-downside, each exp-matching after each step of the procedure is shown.
    The white circle is used to depict the boundary of multiple continuous line segments. 
    }
    \label{fig:thm min inf alg}
\end{figure}
    After the procedure is done, every box contains at most one line segment, by Step~4 of 
    the procedure.
    Also, notice that every step of each procedure cannot result in different costs, as translation is done \emph{inside} the box.
    \begin{observation}\label{obs:normalization-no-cost-difference}
        The normalization procedure from Algorithm~\ref{algo:segment_clipping} does not change the cost of the exp-matching.
    \end{observation}

    \begin{mdframed}
    \textbf{Definition} For \reales{s} $x,y$ and set $\mathcal{B}\subseteq 
    \{b_{i,j}|1\leq i \leq Flen(x),1\leq j \leq Flen(y) \in\mathbb{N} \}$,
    $T_1(\mathcal{B},x,y)$ is the set of resulting exp-matching when the normalization procedure is applied to the exp-matching in $T_0(\mathcal{B},x,y)$.
    \end{mdframed}
    
    For convenience, we denote $T_1(B,p_1,p_2)$ as $T_1$.
    By Observation~\ref{obs:normalization-no-cost-difference}, $\overline{W}(T_0)=\overline{W}(T_1)$. 
    Thus, our theorem is true if there exists an exp-matching 
    with cost $dist(p_1,p_2)$ in $T_1$.
    Notice that any exp-matching in $T_1$ is still covered by $B$ as exp-matching in $T_0$ was covered by $B$.

\textbf{Step 4}: Existence of minimum exp-matching in $T_1$

     Let $B=\{b_{\alpha_1, \beta_1},\cdots b_{\alpha_t, \beta_t}\}$, where $t=|B|\in\mathbb{N}$ 
     and $\alpha_i\le\alpha_{i+1}, \beta_i\le\beta_{i+1}$ for $1\le i \le t$. 
     (Thus, $b_{\alpha_i,\beta_i}$ is $i$-th box area of $B$ from down left to upward right. Recall that $\forall b_{i_0,j_0},b_{i_1,j_1}\in B,
    i_0<i_1 \implies j_0\leq j_1$.)
    Now consider an arbitrary exp-matching $m\in T_1$.
    We let $\overline{x_i}(m)$ as the length of the line segment in $b_{\alpha_i,\beta_i}$ for integer $1\le i \le t$.
    (Recall that each $b\in B$ contains at most one line segment after the normalization procedure.)
    If $b_{\alpha_i,\beta_i}$ does not contain line segment, $\overline{x_i}(m)=0$.

    Now, one can observe $\overline{x_1}(m),\overline{x_2}(m),\cdots, \overline{x_t}(m)$ is unique for
    every exp-matching $m\in T_1$.
    It is because every line segment in every exp-matching in $T_1$ 
    is maximally aligned to the right and up inside predetermined boxes, by the normalization procedure.
    Even more, cost $\overline{W}(m)$ can be expressed as the first order polynomial of $\overline{x_1}(m),\overline{x_2}(m),\cdots, \overline{x_t}(m)$. 
    This implies that our goal is transitioned from finding exp-matching $m$ that minimizes the cost
    to finding valid $\overline{x_1}(m),\overline{x_2}(m),\cdots, \overline{x_t}(m)$, that minimizes the cost. 
    The precise polynomial is shown in Observation~\ref{obs:F-x_1-x_t} and
    the domain is shown in Observation~\ref{obs:condition-F-x_1-x_t} which 
    will be recapped with an example later.

    We define function~$F$ so that $F(x_1,\cdots ,x_t)=\overline{W}(\hat{m})$ where 
    $\hat{m}$ is an exp-matching in $T_1$ that satisfies 
    $(x_1,\cdots ,x_t)=(\overline{x_1}(\hat{m}),\overline{x_2}(\hat{m}),\cdots, \overline{x_t}(\hat{m}))$. 
    Then we obtain Observation~\ref{obs:F-x_1-x_t}.
    \begin{observation}\label{obs:F-x_1-x_t}
    For some $(x_1,\cdots ,x_t)\in \mathbb{R}^t$, if there exists $\hat{m}$ in $T_1$ that satisfies 
    $(x_1,\cdots ,x_t)=(\overline{x_1}(\hat{m}),\overline{x_2}(\hat{m}),\cdots, \overline{x_t}(\hat{m}))$, 
    $F(x_1,\cdots ,x_t)=\overline{W}(\hat{m})$ is represented as follows:
    \begin{align*}
        &F(x_1,\cdots ,x_t)=(\textnormal{Cost of deletion})+(\textnormal{Cost of insertion})+(\textnormal{Cost of substitution})\\
        &=\int_0^{len(p_1)}w_{del}(p_1[x])dx\\
        &-\Big\{w_{del}(P_1(b_{\alpha_1, \beta_1}))\times \frac{x_1}{\sqrt{2}}+w_{del}(P_1(b_{\alpha_2, \beta_2}))\times \frac{x_2}{\sqrt{2}}+\cdots +w_{del}(P_1(b_{\alpha_t, \beta_t}))\times\frac{x_t}{\sqrt{2}} \Big\}\\
        &+\int_0^{len(p_2)}w_{ins}(p_2[x])dx\\
        &-\Big\{w_{ins}(P_2(b_{\alpha_1, \beta_1}))\times \frac{x_1}{\sqrt{2}}+w_{ins}(P_2(b_{\alpha_2, \beta_2}))\times \frac{x_2}{\sqrt{2}}+\cdots +w_{ins}(P_2(b_{\alpha_t, \beta_t}))\times \frac{x_t}{\sqrt{2}}\Big\}\\
        &+\Big\{w_{sub}(P_1(b_{\alpha_1, \beta_1}),P_2(b_{\alpha_1, \beta_1}))\times \frac{x_1}{\sqrt{2}}
        +w_{sub}(P_1(b_{\alpha_2, \beta_2}),P_2(b_{\alpha_2, \beta_2}))\times \frac{x_2}{\sqrt{2}}+\cdots\\
        &~~+w_{sub}(P_1(b_{\alpha_t, \beta_t}),P_2(b_{\alpha_t, \beta_t}))\times \frac{x_t}{\sqrt{2}}\Big\}\\\\
        &=\int_0^{len(p_1)}w_{del}(p_1[x])dx+\int_0^{len(p_2)}w_{ins}(p_2[x])dx\\
        &+\Big\{w_{sub}(P_1(b_{\alpha_1, \beta_1}),P_2(b_{\alpha_1, \beta_1}))-w_{del}(P_1(b_{\alpha_1, \beta_1}))-w_{ins}(P_2(b_{\alpha_1, \beta_1})) \Big\}\times \frac{x_1}{\sqrt{2}}\\
        &+\Big\{w_{sub}(P_1(b_{\alpha_2, \beta_2}),P_2(b_{\alpha_2, \beta_2}))-w_{del}(P_1(b_{\alpha_2, \beta_2}))-w_{ins}(P_2(b_{\alpha_2, \beta_2})) \Big\}\times \frac{x_2}{\sqrt{2}}\\
        &~~~~\vdots\\
        &+\Big\{w_{sub}(P_1(b_{\alpha_t, \beta_t}),P_2(b_{\alpha_t, \beta_t}))-w_{del}(P_1(b_{\alpha_t, \beta_t}))-w_{ins}(P_2(b_{\alpha_t, \beta_t})) \Big\}\times \frac{x_t}{\sqrt{2}}.  
    \end{align*}
    \end{observation}
    $P_1, P_2$ was defined in \textbf{Step~1} of basic notations, as the corresponding character of the box in $p_1$ and $p_2$.
    By triangle inequality, deleting a character and inserting another character have greater or equal cost than direct substitution. 
    Therefore, every coefficient in the equation of Observation~\ref{obs:F-x_1-x_t} is not positive.
    Also, $\int_0^{len(p_1)}w_{del}(p_1[x])dx+\int_0^{len(p_2)}w_{ins}(p_2[x])dx$ is constant.
    Then in Observation~\ref{obs:condition-F-x_1-x_t}, we identify domain of 
    $F$, or the precise condition that $(x_1,\ldots ,x_t)$ should satisfy for existence of corresponding exp-matching~$\hat{m}$ in $T_1$.
    \begin{observation}\label{obs:condition-F-x_1-x_t}
    The tuple~$(x_1, \ldots, x_t)\in \mathbb{R}^t$ satisfies the following conditions if and only if there exists $\hat{m}\in T_1$ that have $(x_1, \ldots, x_t)$ as lengths of their line segments. In other words, $(x_1, \cdots , x_t)=(\overline{x_1}(\hat{m}), \cdots , \overline{x_t}(\hat{m}))$.
    \begin{enumerate}
        \item For every $1\leq i\leq n$, $\sum_{v=1}^{t} \delta(\alpha_v,i)\frac{1}{\sqrt{2}} x_v \leq c_i$.
        \item For every $1\leq j\leq k$, $\sum_{v=1}^{t} \delta(\beta_v,j) \frac{1}{\sqrt{2}} x_v \leq d_j$.
        \item For every $1\leq i\leq t$, $x_i\geq 0$.
    \end{enumerate}
    \end{observation}
    These conditions follow from Condition~3 from Definition~\ref{def:alter-pseudo-edit-dist} and 
    the simple geometric fact that line segments cannot have negative length.
    But such set of $(x_1,\cdots ,x_t)$ is a closed, bounded subset of $\mathbb{R}^t$. 
    Therefore, $(x_1,\cdots ,x_t)$ that minimizes $F(x_1,\cdots ,x_t)$ always exists, proving our theorem. Moreover, this can be extended as a standard form of linear programming problem.
    
    To recap \textbf{Step~4} with an example, consider Figure~\ref{fig:thm min inf}. Depicted exp-matching $E$ lies on $b_{\alpha_1,\beta_1}=b_{1,1},b_{\alpha_2,\beta_2}=b_{1,2},b_{\alpha_3,\beta_3}=b_{1,3}$ and $b_{\alpha_4,\beta_4}=b_{3,3}$.
    $E$ has cost that is 1st order polynomial:
    \begin{align*}
        &F(x_1,x_2,x_3,x_4)=\\
        &~\int_{[0,3.5)-E_X}w_{del}(p_1[x])dx+\int_{[0,3.5)-E_Y}w_{ins}(p_2[y])dy+\frac{1}{\sqrt{2}}\int_{E} w_{sub}(p_1[r_x],p_2[r_y])d\vec{r}\\
        &~=\left\{\left(1.5-\frac{1}{\sqrt{2}}(x_1+x_2+x_3)\right)w_{del}(a)+w_{del}(c)+\left(1-\frac{1}{\sqrt{2}}x_4\right)w_{del}(b)\right\}+\\
        &~~\left\{\left(1-\frac{1}{\sqrt{2}}x_1\right)w_{ins}(c)+\left(1.5-\frac{1}{\sqrt{2}}x_2\right)w_{ins}(b)+\left(1-\frac{1}{\sqrt{2}}(x_3+x_4)\right)w_{ins}(d)\right\}+\\
        &~~\left\{\frac{1}{\sqrt{2}}x_1 w_{sub}(a,c)+\frac{1}{\sqrt{2}}x_2 w_{sub}(a,b)+\frac{1}{\sqrt{2}}x_3 w_{sub}(a,d)+\frac{1}{\sqrt{2}}x_4 w_{sub}(b,d)\right\}\\\\
        &=1.5w_{del}(a)+w_{del}(c)+w_{del}(b)+w_{ins}(c)+1.5w_{ins}(b)+w_{ins}(d)
        \\
        &~+\Big(w_{sub}(a,c)-w_{del}(a)-w_{ins}(c)\Big)\frac{x_1}{\sqrt{2}}\\
        &~+\Big(w_{sub}(a,b)-w_{del}(a)-w_{ins}(b)\Big)\frac{x_2}{\sqrt{2}}\\
        &~+\Big(w_{sub}(b,d)-w_{del}(b)-w_{ins}(d)\Big)\frac{x_3}{\sqrt{2}}.
    \end{align*}
    Also, the precise conditions that $x_1,x_2,x_3$ and $x_4$ should have are, 
    \begin{enumerate}
        \item $\frac{1}{\sqrt{2}}(x_1+x_2+x_3)\leq 1.5$, $0\leq 1$ and $\frac{1}{\sqrt{2}}x_4\leq 1$.
        \item $\frac{1}{\sqrt{2}}x_1\leq 1$, $\frac{1}{\sqrt{2}}x_2\leq 1.5$ and $\frac{1}{\sqrt{2}}(x_3+x_4)\leq 1$.
        \item $x_1,x_2,x_3,x_4\geq 0$.
    \end{enumerate}
    These conditions come from Condition~3 of Definition~\ref{def:alter-pseudo-edit-dist} and 
    the simple geometric fact that length cannot be negative, as argued before.
    For instance, $\frac{1}{\sqrt{2}}(x_1+x_2+x_3)\leq 1.5$ reflects that
    horizontal width of left three line segments in Figure~\ref{fig:thm min inf}
    cannot be greater than 1.5.
    \begin{figure}
        \centering
        \includegraphics[width=0.5\linewidth]{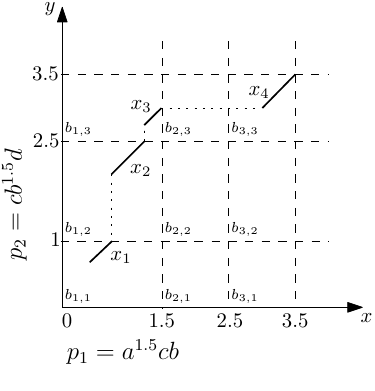}
        \caption{This can be valid exp-matching in $T_1$, if exp-matching in $T_0$ are covered by $B=\{b_{1,1},b_{1,2},b_{1,3},b_{3,3}\}$.
        $x_1,x_2, x_3$ and $x_4$ denote length of each line segment.
        }
        \label{fig:thm min inf}
    \end{figure}
\end{proof}

Proof of Theorem~\ref{thm:inf-min}
admits the existence of an exp-matching that has a minimum cost, thereby allowing us 
to relax Definitions~\ref{def:pseudo-edit-dist} and~\ref{def:alter-pseudo-edit-dist} as
minimum cost, instead of infimum.

While we relaxed the Definitions to more practical way, we now have enough 
foundation to move to our second main point, \textbf{Relation between string edit distance and exp-edit distance}. 

\begin{corollary}\label{cor:string-pseudo-equiv}
When $p_1, p_2\in \Sigma_\mathbb{N}^*$, the string edit distance from $p_1$ to $p_2$ is the same as exp-edit distance.
\end{corollary}

\begin{proof}
    Let $d$ denote the string edit distance from $p_1$ to $p_2$.
    We define string exp-matching, which is an exp-matching with further restriction
    so that all endpoints of the line segments lie on the integer coordinate,
    to represent the exp-matching for string edit distance.
    
    Suppose that Corollary~\ref{cor:string-pseudo-equiv} is false.
    Then $dist(p_1,p_2)<d$, as every string exp-matching is an exp-matching.
    Our strategy is to prove that there is an exp-matching such that its cost is $dist(p_1,p_2)$, and
    is a string exp-matching. This leads to a contradiction, as $d$ is the minimum possible cost
    for string exp-matching.
    
    As our goal to find string exp-matching with minimum cost is similar 
    to the goal of Theorem~\ref{thm:inf-min} to find exp-matching with minimum cost, 
    it is useful to borrow some notations from proof of Theorem~\ref{thm:inf-min}. 
    Let $p_1=\sigma_1^{c_1}\cdots \sigma_n^{c_n}$ and $p_2=\gamma_1^{d_1}\cdots \gamma_k^{d_k}$ at its contraction form where 
    $c_1, \cdots, c_n, d_1, \cdots d_k$ are integers. 
    For $1\leq i \leq n$ and $1\leq j \leq k$, again let $b_{i,j}$ denote the set of points 
    that is at the interior and boundary of the box that is formed by vertical lines 
    $x=\sum_{v=0}^{i-1} c_v$, $x=\sum_{v=0}^{i} c_v$ 
    and horizontal lines $y=\sum_{v=0}^{h-1} d_v$, $y=\sum_{v=0}^{h} d_v$.
    
    For some $B=\{b_{\alpha_1, \beta_1}, \cdots,  b_{\alpha_t, \beta_t} \}$, 
    there exists exp-matching $E\in T_1(B,p_1,p_2)$ with cost $dist(p_1,p_2)$ by Theorem~\ref{thm:inf-min}.
    If $\overline{x_1}(E),\overline{x_2}(E),\cdots ,\overline{x_t}(E)$ are integer multiples of $\sqrt{2}$,
    the normalization procedure from Algorithm~\ref{algo:segment_clipping} ensures that all line segments of $E$
    have their endpoints on integer coordinates.

    \begin{mdframed}
    \textbf{Lemma} If $\overline{x_1}(E),\overline{x_2}(E),\cdots ,\overline{x_t}(E)$ are integer multiples of $\sqrt{2}$, 
    all line segments of $E\in T_1(B,p_1,p_2)$ have their endpoints on integer coordinates.\\
    \textbf{Proof} Proof is by induction from the rightmost line segment.
    For the rightmost line segment with length $\overline{x_t}(E)$, 
    the normalization procedure translates the right endpoint to the vertex of the box which is an integer coordinate,
    as there are no line segments located on the right side of the rightmost
    line segment and thus it cannot be blocked by Condition~3 of Definition~\ref{def:alter-pseudo-edit-dist}.
    Then, the left endpoint is also located at the integer coordinate as the 
    length of the line segment $\overline{x_t}(E)$ is some multiple of $\sqrt{2}$.
    
    For the inductive step, assume the line segment $l_{i+1}$ with length $\overline{x_{i+1}}(E)$ have their endpoints located at the integer coordinate.
    For the line segment $l_{i}$ with length $\overline{x_{i}}(E)$, as the left endpoint of $l_{i+1}$ is at the integer coordinate,
    regardless of whether Steps~2 and~3 of the normalization procedure from Algorithm~\ref{algo:segment_clipping}
    is stopped by Condition~3 of Definition~\ref{def:alter-pseudo-edit-dist} 
    or the side of the box, the right endpoint of $l_i$ is at the integer coordinate. As $\overline{x_i}(E)$ is a multiple of $\sqrt{2}$, the left endpoint is also at the integer coordinate. \qed
    \end{mdframed}
    This confirms that $E$ is a string exp-matching.
    We introduce new variables~$(x_1',x_2',\cdots ,x_t') \allowbreak
    = \allowbreak (\frac{1}{\sqrt{2}}x_1,\allowbreak\frac{1}{\sqrt{2}}x_2,\cdots ,\frac{1}{\sqrt{2}}x_t)$,
    which are in the domain specified from Observation~\ref{obs:condition-F-x_1-x_t}.
    We show $\overline{x_1}(E),\overline{x_2}(E),\cdots ,\overline{x_t}(E)$ are integer multiples of $\sqrt{2}$
    using  $(\overline{x_1}(E),\overline{x_2}(E),\cdots ,\overline{x_t}(E)) = \argmin_{(x_1,\cdots ,x_t)}F(x_1,\cdots ,x_t)$. 
    If we recap Observation~\ref{obs:F-x_1-x_t} and the conditions that should be satisfied, they are:
    \begin{align*}
        &F(x_1,\cdots ,x_t)=\\
        &\int_0^{len(p_1)}w_{del}(p_1[x])dx+\int_0^{len(p_2)}w_{ins}(p_2[x])dx\\
        &+\Big\{w_{sub}(P_1(b_{\alpha_1, \beta_1}),P_2(b_{\alpha_1, \beta_1}))-w_{del}(P_1(b_{\alpha_1, \beta_1}))-w_{ins}(P_2(b_{\alpha_1, \beta_1})) \Big\}\times x_1'\\
        &+\Big\{w_{sub}(P_1(b_{\alpha_2, \beta_2}),P_2(b_{\alpha_2, \beta_2}))-w_{del}(P_1(b_{\alpha_2, \beta_2}))-w_{ins}(P_2(b_{\alpha_2, \beta_2})) \Big\}\times x_2'\\
        &\vdots\\
        &+\Big\{w_{sub}(P_1(b_{\alpha_t, \beta_t}),P_2(b_{\alpha_t, \beta_t}))-w_{del}(P_1(b_{\alpha_t, \beta_t}))-w_{ins}(P_2(b_{\alpha_t, \beta_t})) \Big\}\times x_t'
    \end{align*}
    and
    \begin{enumerate}
        \item For every $1\leq i\leq n$, $\sum_{v=1}^{t} \delta(\alpha_v,i) x_v' \leq c_i$.
        \item For every $1\leq j\leq k$, $\sum_{v=1}^{t} \delta(\beta_v,j)  x_v' \leq d_j$.
        \item For every $1\leq i\leq t$, $x_i'\geq 0$.
    \end{enumerate}
    We are going to find $(x_1',\cdots ,x_t')$ that let cost $F(x_1,\cdots ,x_t)$ to be minimized in a given condition, and show they are integers, which is 
    equivalent to showing $\overline{x_1}(E),\overline{x_2}(E),\cdots ,\overline{x_t}(E)$ are integer multiples of $\sqrt{2}$.
    Let $\textbf{x}=(x_1',\cdots ,x_t')^\textnormal{T}$.
    Let $\textbf{A}$ be a $(n+k)\times t$ size 0-1 matrix defined as follows:
    \begin{equation*}
    [\textbf{A}]_{i,j}=
        \begin{cases}
            1 &\textnormal{ if } i\leq n,\alpha_j=i\\
            0 &\textnormal{ if } i\leq n, \alpha_j\not =i\\
            1 &\textnormal{ if } i> n, \beta_j=i-n\\
            0 &\textnormal{ if } i> n, \beta_j\not=i-n\\
        \end{cases}.
    \end{equation*}
    Let $\textbf{b}=(c_1,\cdots ,c_n,d_1,\cdots ,d_k)^\textnormal{T}$. 
    Using those definitions, we can give an alternative problem setting to find $x_1',\cdots ,x_t'$, 
    that satisfies  $F(x_1,\cdots ,x_t)=dist(p_1,p_2)$. That is,
    \begin{align*}
        \textnormal{Minimize } F(x_1,\cdots ,x_t)\\
        \textnormal{Subject to } \textbf{A}\textbf{x}\leq \textbf{b}, \textbf{x}\geq 0.
    \end{align*}
    However, this is the same as 
    \begin{align*}
        \textnormal{Maximize }& \sum_{i=1}^t \left(w_{del}(P_1(b_{\alpha_i, \beta_i}))+w_{ins}(P_2(b_{\alpha_i, \beta_i})) -w_{sub}(P_1(b_{\alpha_i, \beta_i}),P_2(b_{\alpha_i, \beta_i})) \right)
        \times x_i'\\
        \textnormal{Subject to }& \textbf{A}\textbf{x}\leq \textbf{b}, \textbf{x}\geq 0.
    \end{align*}
    Notice that $\sum_{i=1}^t (w_{del}(P_1(b_{\alpha_i, \beta_i}))+w_{ins}(P_2(b_{\alpha_i, \beta_i})) -w_{sub}(P_1(b_{\alpha_i, \beta_i}),P_2(b_{\alpha_i, \beta_i})) )\times x_i'$ 
    is expanded equation of $-F(x_1,\cdots ,x_t)+\int_0^{len(p_1)}w_{del}(p_1[x])dx+\int_0^{len(p_2)}w_{ins}(p_2[x])dx$. 
    Coefficients $w_{del}(P_1(b_{\alpha_i, \beta_i}))+w_{ins}(P_2(b_{\alpha_i, \beta_i})) -w_{sub}(P_1(b_{\alpha_i, \beta_i}),P_2(b_{\alpha_i, \beta_i}))$
    are positive by triangle inequality, so this is a standard linear programming problem setting. 
    If we demonstrate that the solution~$\textbf{x}$ is an integer, it confirms that $E$ qualifies as string exp-matching.
    The known sufficient condition for $\textbf{x}$ to be an integer solution is when $\textbf{A}$ is totally unimodular. 
    Again, a sufficient condition for $\textbf{A}$ to be totally unimodular is when $\textbf{A}$ is the incidence matrix of a bipartite matching graph, and that is true.
    Consider $j$th column of $\textbf{A}$. By definition of $\textbf{A}$, it should have two entries of 1:
    One at $\alpha_j$-th row, and the other at $\beta_j+n$-th row. 
    Therefore, if we interpret the first $n$ rows as one 
    part of the bipartite graph, and the following $k$ rows as the other part, it is clear that two parts are bipartite and
    thus $\textbf{A}$ is the incidence matrix of a bipartite matching graph.
\end{proof}

We move to our third point of 
\textbf{Identifying how exp-edit distance changes 
when exponents of $p$ and $q$ are multiplied by $C$}
by employing geometrical viewpoint on exp-matching and 
existence of minimum cost exp-matching from Theorem~\ref{thm:inf-min}.
Theorem~\ref{thm:similar} addresses this in a generalized way, 
assuming an arbitrary positive real number $k$ is multiplied to
the exponents.
\begin{theorem}\label{thm:similar}
    Given an alphabet $\Sigma$, let
    $p_1=\sigma_1^{c_1}\sigma_2^{c_2}\cdots \sigma_n^{c_n}$,
    $p_2=\gamma_1^{d_1}\gamma_2^{d_2}\cdots \gamma_m^{d_m}$,
    $p_1'=\sigma_1^{kc_1}\sigma_2^{kc_2}\cdots \sigma_n^{kc_n}$ and 
    $p_2'=\gamma_1^{kd_1}\gamma_2^{kd_2}\cdots \gamma_m^{kd_m}$ 
    where
    $n,m\in\mathbb{N}$, $k\in\mathbb{R}^+$ and $\sigma_i,\gamma_j\in\Sigma, c_i,d_j\in \mathbb{R}^+$ for $1\leq i \leq n, 1\leq j \leq m \in \mathbb{N}$.
    Then, $dist(p_1',p_2')=k\times dist(p_1,p_2)$.
\end{theorem}
\begin{proof}
    Let $E$ be one of the exp-matching from $p_1$ to $p_2$ with the minimum cost,
    and let $E'$ be one of the exp-matching from $p_1'$ to $p_2'$ with the minimum cost.
    Then, we claim for some $E$ and $E'$, they are \emph{similar} by the ratio of $1:k$.
    Here, \emph{similar} is a geometry terminology~\cite{Piaget48}; One shape can be obtained by scaling up
    the other shape.
    Assume that the claim is false. Notice that an exp-matching gained by scaling 
    $E$ with the ratio $k$ is a valid exp-matching from $p_1'$ to $p_2'$.
    This exp-matching has a cost of $k\times dist(p_1,p_2)$, and it must be greater than
    the cost of $E'$ as $E'$ is the exp-matching from $p_1'$ to $p_2'$ with the minimum cost.
    Therefore, $k\times dist(p_1,p_2)>dist(p_1',p_2')$.
    Conversely, consider an exp-matching gained by scaling $E'$ by the ratio $1/k$.
    This is a valid exp-matching from $p_1$ to $p_2$, and
    has a cost $\frac{1}{k}\times dist(p_1',p_2')$. 
    As $E$ is the exp-matching from $p_1$ to $p_2$ that have minimum cost,
    $\frac{1}{k}\times dist(p_1',p_2')>dist(p_1,p_2)$.
    This leads to a contradiction to our previous observation.
    Therefore, our claim is true, and Theorem~\ref{thm:similar} is a direct consequence of
    the claim.
\end{proof}

\begin{theorem}\label{thm:algo-correct}
    Algorithm~\ref{algo:exp-edit distance calculation} is correct.
\end{theorem}
\begin{proof}
    By Corollary~\ref{cor:string-pseudo-equiv} 
    and Theorem~\ref{thm:similar}, 
    $d= dist(w_1,w_2) = C\times dist(p,q)$. 
    Therefore, $dist(p,q)=d/C$.
\end{proof}
\subsection{Time Complexity Analysis}\label{ssec:time-complexity}
Let $n$ (respectively, $m$) be the number of factors in 
a given representation of \rates{} $p$.
For cases where one of the given \rates{s} is $\lambda$, 
it takes $O(\max(n,m))$ time and terminates at Step 0.
In Step 1, converting $p$ and $q$ to the contraction form takes $O(n+m)$.
$C$ can be calculated by multiplying every denominator, which takes 
$O(n+m)$ time \emph{assuming multiplication takes $O(1)$ time}.
$w_1$ (respectively, $w_2$) is computed in $O(n)$ (respectively, $O(m)$) time.

Most string edit distance algorithm takes exponential time 
in Step 2, as the lengths of $w_1$ and $w_2$ are exponential to the size of the input.
However, if $w_{ins}(a)=w_{del}(a)=w_{sub}(a,b)=1$ for $a\neq{}b\in\Sigma$,
an algorithm that works on RLE string takes
$O(mn\log{(mn)})$ time~\cite{CliffordGKMU19}.
Finally, Step 3 takes $O(1)$.

In total, if the cost of every string edit operation is 1, 
Algorithm~\ref{algo:exp-edit distance calculation} takes $O(mn\log{mn})$ time.
Otherwise, it takes exponential time.

\section{Properties of exp-edit distance}\label{ssec:investigate-editdist-property}
We now move to study further properties of exp-edit distance to 
enrich our understanding about exp-edit distance.
In particular, we highlight similarities with string edit distance, 
fortifying our claim that exp-edit distance is a natural extension of
string edit distance.

These are well-known properties of string edit distance. 
\begin{itemize}
    \item For strings~$w,u,v\in\Sigma_{\mathbb{N}}^*$ and string edit distance $d(\cdot ,\cdot)$, $d(wu,wv)=d(u,v)$ and $d(uw,vw)=d(u,v).$
    \item For the case of Levenshtein edit distance where cost of insertion, deletion and substitution is 1, $d(\cdot ,\cdot)$ satisfies the metric axioms.
\end{itemize}
We show that these properties hold even for exp-edit distance.

\begin{theorem}\label{thm:prefix}
For \reales{s}~$w, u, v$,
\[
dist(wu,wv)=dist(u,v) \text{ and } dist(uw,vw)=dist(u,v).
\]
\end{theorem}
\begin{proof}
    $dist(wu, wv)$ is not larger than $dist(u,v)$.
    This is because given an arbitrary exp-edit sequence~$\{r_i \}_{0\leq i \leq n}$ from $u$ to $v$, 
    an exp-edit sequence~$\{wr_i \}_{0\leq i \leq n}$ from $wu$ to $wv$ have the same cost.
    Let us assume that $dist(wu,wv)<dist(u,v)$ to assert a proof by construction.
    Let $E$ be the exp-matching of the cost $dist(wu,wv)$.
    Then, consider the partitions~$L$ and $R$ of $E$ that are at the left and right sides of the vertical line~$x=len(w)$, respectively.
    Let $k=\sup\{y|(x,y)\in L\}$.
    Then not only every $y$ coordinate of elements of $L$ is smaller than or equal to $k$, 
    but also every $y$ coordinate of elements of $R$ is greater than or equal to $k$
    by Condition~4 of an exp-matching. 

    Then, one can construct an exp-edit sequence divided into two parts, part~(1)~and~(2),
    using the same way with Theorem~\ref{thm:alter-def-equiv-lower}:
    edit operations corresponding to the line segments of 
    $L\subseteq [len(w),len(wu))\times [k,len(wv))$ occur in the part~(1) and
    edit operations corresponding to the line segments of $R\subseteq[0,len(w))\times [0,k)$ occur in the part~(2).
    
    This exp-edit sequence should be denoted as follows:
    $wu\overbrace{\Rightarrow\cdots \Rightarrow}^{part~(1)}wu'\overbrace{\Rightarrow\cdots \Rightarrow}^{part~(2)} w'u'=wv$.
    The cost of part~(1) is the cost of changing $u$ to $u'$.
    As $w'u'=wv$, $w'$ is a prefix of $w$ or $w$ is a prefix of $w'$.
    Assume that $w$ is a prefix of $w'$, then $w'=ww''$.
    Then, $w'u' = ww''u'$ and $v=w''u'$.
    
    The cost of part~(2) is the cost of inserting $w''$ after $w$.
    Thus, the total cost of the exp-edit sequence is changing $u$ to $w''u'$ which is $v$.
    It is straightforward that the total cost of the exp-edit sequence is $dist(u,v)$.
    
    When $w'$ is a prefix of $w$, $w=w'w''$, then $u'=w''v$.
    Then, similarly, the cost of part~(2) is the cost of deleting the whole~$w''$ from $w$.
    Therefore, the total cost of the exp-edit sequence is the cost of changing $wu=w'w''u$ into $wu'=w'w''w''v$, and then deleting $w''$ from $wu'$.
    If this cost is smaller than $dist(u,v)$,
    it is a contradiction as $u\Rightarrow\cdots\Rightarrow u'=w''v \Rightarrow\cdots\Rightarrow v$ is also valid exp-edit sequence from $u$ to $v$ and have same cost.
    For $dist(uw,vw)=dist(u,v)$, the symmetric argument proves it.
\end{proof}

\begin{corollary}\label{cor:prefix-suffix-together}
For \reales{s}~$x,y, u,v$, $dist(xuy,xvy)=dist(u,v)$.
\end{corollary}
\begin{proof}
    This is straightforward from
    $dist(xuy,xvy)=dist(uy,vy)=dist(u,v)$,
    which is a direct consequence of Theorem~\ref{thm:prefix}.
\end{proof}

\begin{proposition}\label{prop:metric-axiom}
    If $w_{ins}(a)=w_{del}(a)=1$ and $w_{sub}(a,b)=w_{sub}(b,a)=1$ for arbitrary $a\not = b \in \Sigma$,
    and $p,q,r \in\Sigma_{\mathbb{R}^{+}}^* $,
    the following properties known as metric axioms hold:
    \begin{align}
        &dist(p,p)=0\label{property:1}\\
        &\textnormal{If } p\not = q, dist(p,q)>0\label{property:2}\\
        &dist(p,q)=dist(q,p)\label{property:3}\\
        &dist(p,r)\leq dist(p,q)+dist(q,r)\label{property:4}
    \end{align}
\end{proposition}
\begin{proof}
    Properties~(\ref{property:1}) and (\ref{property:2}) are trivial by definition.
    For Property~(\ref{property:3}), as $w_{ins}(a)=w_{del}(a)$ and $w_{sub}(a,b)=w_{sub}(b,a)$ for
    every $a\not=b\in \Sigma$, every exp-edit sequence from $p$ to $q$ corresponds to a symmetric,
    same cost exp-edit sequence from $q$ to $p$.
    Therefore, Property~(\ref{property:3}) is also easy to know. 
    Assume Property~(\ref{property:4}) isn't true for some \reales{s}~$p,q,r$.
    Then, we have $dist(p,r)> dist(p,q)+dist(q,r)$.
    Consider an exp-edit sequence from $p$ to $r$, gained by the first edit $p$ to be $q$, and edit $q$ to be $r$.
    The minimum possible cost of such an exp-edit sequence is $dist(p,q)+dist(q,r)$.
    However, this leads to a contradiction because $dist(p,r)$ should be the minimum cost of the exp-edit sequence
    from $p$ to $r$ but we assumed $dist(p,r)>dist(p,q)+dist(q,r)$.
\end{proof}

String edit distance and exp-edit distance is defined differently, 
in a sense that string edit sequence 
is consisted of $\mathbb{N}$-exponent-strings.
Thus, Corollary~\ref{cor:string-pseudo-equiv}, which gives that exp-edit distance of $\mathbb{N}$-exponent-strings is same with string edit distance can be alternatively stated without proof.
\begin{proposition}\label{prop:alternative-statement-string-pseudo-equiv}
    For $p_1,p_2\in\Sigma_{\mathbb{N}}^*$, there exists exp-edit sequence from $p_1$ to $p_2$ consisted of $\mathbb{N}$-exponent-strings.
\end{proposition}
Based on Theorem~\ref{thm:similar}, deriving a rational analogous statement
of Proposition~\ref{prop:alternative-statement-string-pseudo-equiv} is straightforward.

\begin{theorem}\label{thm:rational-pseudo-equiv}
    For arbitrary \rates{s}~$p_1,p_2$ 
    there exists a minimum cost exp-edit sequence from $p_1$ to $p_2$ 
    consisted of only \rates{s}.
\end{theorem}
\begin{proof}
    Let $k$ be the common denominator of every exponent of factors in the contraction form of 
    $p_1$ and $p_2$.
    If $p_1=\sigma_1^{c_1}\sigma_2^{c_2}\cdots \sigma_n^{c_n}$ and
    $p_2=\gamma_1^{d_1}\gamma_2^{d_2}\cdots \gamma_m^{d_m}$, then
    $p_1'=\sigma_1^{kc_1}\sigma_2^{kc_2}\cdots \sigma_n^{kc_n}$ and 
    $p_2'=\gamma_1^{kd_1}\gamma_2^{kd_2}\cdots \gamma_m^{kd_m}$
    are \emph{strings}, as every exponent is an integer in the contraction form.
    Thus, by Theorem~\ref{thm:similar},
    $dist(p_1,p_2)=k\times dist(p_1',p_2')$.
    As $p_1'$ and $p_2'$ are strings, by Proposition~\ref{prop:alternative-statement-string-pseudo-equiv},
    let $K'$ be the minimum cost exp-edit sequence from $p_1'$ to $p_2'$ consisted of only strings.
    For every term~$s$ of $K'$, consider multiplying $\frac{1}{k}$ to every
    exponent of $s$ to get a new exp-edit sequence $K$.
    As this is exactly reverse way of construction of $p_1'$ and $p_2'$ from $p_1$ and $p_2$,
    $K$ is a valid exp-edit sequence from $p_1$ to $p_2$, with cost $\frac{1}{k} dist(p_1',p_2')=dist(p_1,p_2)$.
    $K$ consists of terms that are only \rates{s},
    as multiplying $\frac{1}{k}$ to some integer always results in a rational number.
\end{proof}
Based on the insights from Corollary~\ref{cor:string-pseudo-equiv} and Theorem~\ref{thm:rational-pseudo-equiv},
we can establish Corollary~\ref{cor:hierachty-edit-distance} without further proof.
\begin{corollary}\label{cor:hierachty-edit-distance}
    Let $dist_{\mathbb{Q}}:\Sigma_{\mathbb{Q}^+}^* \times \Sigma_{\mathbb{Q}^+}^* \rightarrow \mathbb{R}_{\geq 0}$
    be the minimum cost exp-edit sequence consisted of only \rates{s} and $dist_{\mathbb{N}}:\Sigma_{\mathbb{N}}^* \times \Sigma_{\mathbb{N}}^* \rightarrow \mathbb{R}_{\geq 0}$ be the string edit distance.
    Then, \[
        dist_{\mathbb{N}} \subseteq dist_{\mathbb{Q}} \subseteq dist \subseteq \Sigma_{\mathbb{R}^+}^* \times \Sigma_{\mathbb{R}^+}^* \times \mathbb{R}_{\geq 0}.
    \]
\end{corollary}

Corollary~\ref{cor:hierachty-edit-distance} illustrates that the hierarchy of edit distances
from natural to real numbers broadens the scope of applicable sequences,
reflecting an increase in complexity and flexibility in \es{s}.

\section*{Acknowledgements}
I would like to thank Joonghyuk Hahn, Yo-Sub Han, and Kai Salomaa for their valuable 
discussions and insightful feedback, which significantly contributed to this research.
\bibliographystyle{plain}
\bibliography{reference}

\begin{thebibliography}{10}

\bibitem{bruce1989report}
G{\"o}sta Bruce.
\newblock Report from the ipa working group on supra-segmental categories.
\newblock {\em Working papers/Lund University, Department of Linguistics and Phonetics}, 35:25--40, 1989.

\bibitem{CliffordGKMU19}
Rapha{\"{e}}l Clifford, Pawel Gawrychowski, Tomasz Kociumaka, Daniel~P. Martin, and Przemyslaw Uznanski.
\newblock {RLE} edit distance in near optimal time.
\newblock In {\em 44th International Symposium on Mathematical Foundations of Computer Science}, volume 138 of {\em LIPIcs}, pages 66:1--66:13, 2019.

\bibitem{Docherty92}
Gerard~J Docherty.
\newblock {\em The timing of voicing in British English obstruents}.
\newblock Walter de Gruyter, 1992.

\bibitem{JacksonA77}
Paul~H Jackson and Christian~C Agunwamba.
\newblock Lower bounds for the reliability of the total score on a test composed of non-homogeneous items: I: Algebraic lower bounds.
\newblock {\em Psychometrika}, 42:567--578, 1977.

\bibitem{Johnson04}
Keith Johnson.
\newblock {Aligning phonetic transcriptions with their citation forms}.
\newblock {\em Acoustics Research Letters Online}, 5(2):19--24, 2004.

\bibitem{Keating88}
Patricia~A Keating.
\newblock Underspecification in phonetics.
\newblock {\em Phonology}, 5(2):275--292, 1988.

\bibitem{KessensS04}
Judith~M Kessens and Helmer Strik.
\newblock On automatic phonetic transcription quality: lower word error rates do not guarantee better transcriptions.
\newblock {\em Computer Speech \& Language}, 18(2):123--141, 2004.

\bibitem{KluenderDW88}
Keith~R Kluender, Randy~L Diehl, and Beverly~A Wright.
\newblock Vowel-length differences before voiced and voiceless consonants: An auditory explanation.
\newblock {\em Journal of phonetics}, 16(2):153--169, 1988.

\bibitem{LadefogedM90}
Peter Ladefoged and Ian Maddieson.
\newblock Vowels of the world’s languages.
\newblock {\em Journal of Phonetics}, 18(2):93--122, 1990.

\bibitem{Lauf89}
Raphaela Lauf.
\newblock International phonetic association 1989 kiel convention: The revision of the international phonetic alphabet kiel.
\newblock {\em Zeitschrift für Dialektologie und Linguistik}, 56(3):328--336, 1989.

\bibitem{LeLP17}
Duc Le, Keli Licata, and Emily~Mower Provost.
\newblock Automatic paraphasia detection from aphasic speech: {A} preliminary study.
\newblock In {\em Interspeech 2017, 18th Annual Conference of the International Speech Communication Association, Stockholm, Sweden, August 20-24, 2017}, pages 294--298. {ISCA}, 2017.

\bibitem{LevinsonLM90}
S.E. Levinson, A.~Ljolje, and L.G. Miller.
\newblock Continuous speech recognition from a phonetic transcription.
\newblock In {\em International Conference on Acoustics, Speech, and Signal Processing}, volume~1, pages 93--96, 1990.

\bibitem{Ohman66}
Sven~EG {\"O}hman.
\newblock Coarticulation in vcv utterances: Spectrographic measurements.
\newblock {\em The Journal of the Acoustical Society of America}, 39(1):151--168, 1966.

\bibitem{OconnellK94}
Daniel~C O’connell and Sabine Kowal.
\newblock Some current transcription systems for spoken discourse: A critical analysis.
\newblock {\em Pragmatics. Quarterly Publication of the International Pragmatics Association (IPrA)}, 4(1):81--107, 1994.

\bibitem{PamisettyS23}
Giridhar Pamisetty and K~Sri Rama~Murty.
\newblock Prosody-tts: An end-to-end speech synthesis system with prosody control.
\newblock {\em Circuits, Systems, and Signal Processing}, 42(1):361--384, 2023.

\bibitem{Piaget48}
Jean Piaget and Barbel Inhelder.
\newblock La repr{\'e}sentation de l'espace chez l'enfant.
\newblock In {\em La repr{\'e}sentation de l'espace chez l'enfant}, pages 581--581. Presses Universitaires de France, 1948.

\bibitem{Pitman1848}
Isaac Pitman and Alexander~John Ellis.
\newblock {\em Phonotypic Journal}, volume~7.
\newblock F. Pitman, 1848.

\bibitem{quek00}
Francis~KH Quek.
\newblock An algorithm for the rapid computation of boundaries of run-length encoded regions.
\newblock {\em Pattern recognition}, 33(10):1637--1649, 2000.

\bibitem{SaraclarK00}
Murat Saraclar and Sanjeev Khudanpur.
\newblock Pronunciation ambiguity vs. pronunciation variability in speech recognition.
\newblock In {\em Proceedings of the IEEE International Conference on Acoustics, Speech, and Signal Processing}, volume~3, pages 1679--1682, 2000.

\bibitem{Smith19}
Michael Smith, Kevin~T Cunningham, and Katarina~L Haley.
\newblock Automating error frequency analysis via the phonemic edit distance ratio.
\newblock {\em Journal of Speech, Language, and Hearing Research}, 62(6):1719--1723, 2019.

\bibitem{Thorsen87}
Nina~Gr{\o}nnum Thorsen.
\newblock Suprasegmental transcription.
\newblock {\em Annual Report of the Institute of Phonetics University of Copenhagen}, 21:1--27, 1987.

\bibitem{WesterKCS01}
Mirjam Wester, Judith~M Kessens, Catia Cucchiarini, and Helmer Strik.
\newblock Obtaining phonetic transcriptions: A comparison between expert listeners and a continuous speech recognizer.
\newblock {\em Language and Speech}, 44(3):377--403, 2001.

\bibitem{YinZC21}
Ningjie Yin, Hua Zhang, and Yajie Chen.
\newblock Error analysis of the english vowels based on experimental phonetics study.
\newblock {\em International Journal of Education and Management}, page 170, 2021.

\end{thebibliography}

\clearpage
\end{document}